\documentclass[11pt]{article}
\pdfoutput=1
\usepackage{amsmath,amssymb,amsthm,amsxtra,overpic,bbm,bm,epsfig}
\usepackage{ulem,subfigure}

\def\thefootnote{\fnsymbol{footnote}} 
\allowdisplaybreaks

\begin{document}

\begin{flushright} 
IPPP/16/69  \\
\end{flushright} 

\vspace{0.2cm}

\begin{center}
{\Large\bf Flavon-induced connections between lepton flavour mixing and charged lepton flavour violation processes}
\end{center}

\vspace{0.2cm}

\begin{center}
{\bf Silvia Pascoli}\footnote{E-mail: silvia.pascoli@durham.ac.uk}
and 
{\bf Ye-Ling Zhou}\footnote{E-mail: ye-ling.zhou@durham.ac.uk}
\\
{Institute for Particle Physics Phenomenology, Department of Physics, \\ Durham University, Durham DH1 3LE, United Kingdom} 
\end{center}

\vspace{1.5cm}

\begin{abstract}

In leptonic flavour models with discrete flavour symmetries, couplings between flavons and leptons can result in special flavour structures after they gain vacuum expectation values. At the same time, they can also contribute to the other lepton-flavour-violating processes. We study the flavon-induced LFV 3-body charged lepton decays and radiative decays and we take as example the $A_4$ discrete symmetry. In $A_4$ models, a $Z_3$ residual symmetry roughly holds in the charged lepton sector for the realisation of tri-bimaximal mixing at leading order. The only processes allowed by this symmetry are $\tau^-\to \mu^+ e^- e^-,  e^+ \mu^- \mu^-$, and the other 3-body and all radiative decays are suppressed by small $Z_3$-breaking effects. 
These processes also depend on the representation the flavon is in, whether pseudo-real (case i) or complex (case ii).  
We calculate the decay rates for all processes for each case and derive their strong connection with lepton flavour mixing. 
In case i, sum rules for the branching ratios of these processes are obtained, with typical examples $\text{Br}(\tau^-\to \mu^+ e^- e^-) \approx \text{Br}(\tau^-\to  e^+ \mu^- \mu^-)$ and $\text{Br}(\tau^-\to e^- \gamma) \approx \text{Br}(\tau^-\to  \mu^- \gamma)$. In case ii, we observe that the mixing between two $Z_3$-covariant flavons plays an important role.  
All processes are suppressed by charged lepton masses and current experimental constraints allow the electroweak scale and the flavon masses to be around hundreds of GeV.  
Our discussion can be generalised in other flavour models with different flavour symmetries.

\end{abstract}

\begin{flushleft}
\hspace{0.8cm} PACS number(s): 14.60.Pq, 11.30.Hv, 12.60.Fr, 13.35.-r \\
\hspace{0.8cm} Keywords: Lepton flavour mixing, cross couplings, flavour symmetry, electroweak scale,\\ 
\hspace{2.7cm} charged lepton flavour violation
\end{flushleft}

\def\thefootnote{\arabic{footnote}}
\setcounter{footnote}{0}

\newpage

\tableofcontents

\section{Introduction}

A series of solar \cite{solar}, atmospheric \cite{atmospheric}, accelerator \cite{accelerator} and reactor \cite{reactor} neutrino oscillation experiments have proven that neutrinos have masses and mix.  The mixing is described by the so-called Pontecorvo-Maki-Nakagawa-Sakata (PMNS) matrix \cite{PMNS}, which is parametrised by \cite{PDG}
\begin{eqnarray}
U_\text{PMNS}= 
\begin{pmatrix}
c^{}_{12} c^{}_{13} & s^{}_{12} c^{}_{13} & s^{}_{13} e^{-i
\delta} \\ 
-s^{}_{12} c^{}_{23} - c^{}_{12} s^{}_{13} s^{}_{23}
e^{i \delta} & c^{}_{12} c^{}_{23} - s^{}_{12} s^{}_{13}
s^{}_{23} e^{i \delta} & c^{}_{13} s^{}_{23} \\ 
s^{}_{12}
s^{}_{23} - c^{}_{12} s^{}_{13} c^{}_{23} e^{i \delta} &
-c^{}_{12} s^{}_{23} - s^{}_{12} s^{}_{13} c^{}_{23} e^{i
\delta} & c^{}_{13} c^{}_{23} 
\end{pmatrix} 
\begin{pmatrix}
1 & 0 & 0 \\ 
0 & e^{i\alpha_{21}/2} & 0 \\
0 & 0 & e^{i\alpha_{31}/2} 
\end{pmatrix} 
\; ,
\label{eq:PMNS}
\end{eqnarray}
in which $c_{ij}\equiv\cos\theta_{ij}$ and $s_{ij}\equiv\sin\theta_{ij}$. The three mixing angles have been measured to a good accuracy. Their current best-fit and $\pm1\sigma$ values from a global analysis of the available data \cite{globalfit} are given by 
\begin{eqnarray}
\sin^2\theta_{12} = 0.308^{+0.013}_{-0.012}\,,\;\;
\sin^2\theta_{23} = 0.574^{+0.026}_{-0.144}\,(0.579^{+0.022}_{-0.029})\,,\;\;
\sin^2\theta_{13} = 0.0217^{+0.0013}_{-0.0010}\,(0.0221^{+0.0010}_{-0.0010})
\label{eq:1sigma} 
\end{eqnarray} 
for the normal (inverted) ordering of neutrino masses,  $m_1<m_3$ ($m_1>m_3$). 
We notice that the atmospheric angle $\theta_{23}$ and solar angle $\theta_{12}$ are rather large, with $\theta_{23}$ possibly being maximal, and the reactor angle $\theta_{13}$ takes a value around 0.1, $\theta_{13}\sim 9^\circ$. 
   
The origin of this distinct mixing structure remains unexplained. Discrete flavour symmetries have been widely used to address these questions. It is assumed that at some high energy scale, there exists an underlying discrete flavour symmetry, $G_\text{f}$, which unifies the three flavours together. The tetrahedral group $A_4$ \cite{Ma:2001dn}, which is the smallest group containing 3-dimensional irreducible representations, is the most famous example of this type. There are other commonly-used groups, such as $S_4$ \cite{S4}, $A_5$ \cite{A5}, $\Delta(48)$ \cite{Delta48}, and $\Delta(96)$ \cite{TFH}. At a lower energy scale, the flavour symmetry is broken, leading to nontrivial flavour mixing. 
Most models are built in the framework of the so-called ``direct'' or ``semi-direct'' approaches \cite{King:2013eh}. In these cases, different residual symmetries, $G_l$ and $G_\nu$, subgroups of $G_\text{f}$, are preserved in the charged lepton and neutrino sectors, respectively after the whole flavour symmetry $G_\text{f}$ breaking. By choosing different $G_l$ and $G_\nu$, special flavour structures arise. In the direct approach, the mixing matrix is fully determined by $G_l$ and $G_\nu$ up to Majorana phases and column or row permutations of the PMNS matrix. In the semi-direct approach, $G_l$ and $G_\nu$ cannot fully determine flavour mixing and an accidental symmetry is necessary. For instance, in models based on $A_4$, the tri-bimaximal (TBM) mixing pattern \cite{TBM}, which predicts $s_{12}=1/\sqrt{3}$, $s_{23}=1/\sqrt{2}$ and $s_{13}=0$, can be realised in the semi-direct approach \cite{Altarelli:2005yp, Altarelli:2005yx, A4_TBM}. The residual symmetries are chosen to be $G_l=Z_3$ and $G_\nu=Z_2$, while an additional $Z_2'$ symmetry, not belonging to $A_4$, arises accidentally in the neutrino sector.   
In models of $S_4$, the residual $Z_3, Z_2, Z_2'$ all belong to $S_4$, so TBM can be obtained in the  direct approach \cite{Lam:2008sh, S4_TBM}. $S_4$ can predict another mixing pattern, the bimaximal mixing one ($s_{12}=s_{23}=1/\sqrt{2}$, $s_{13}=0$) \cite{BM} by choosing $G_l=Z_4$ \cite{S4_BM_direct}. Some other mixing patterns can be arranged using larger groups in the direct approach, for instance, the golden-ratio mixing ($s_{12}=\sqrt{2}/\sqrt{5+\sqrt{5}}$, $s_{23}=1/\sqrt{2}$, $s_{13}=0$) \cite{GR} predicted by $A_5$ \cite{A5_GR_direct, Varzielas:2013hga} and the Toorop-Feruglio-Hagedorn mixing ($s_{12}=s_{23}=\sqrt{2}/\sqrt{4+\sqrt{3}}$, $s_{13}=1/(3+\sqrt{3})$) predicted by $\Delta(96)$ \cite{TFH}. It should be noted that the predicted values of $\theta_{13}$ in all these mixing patterns is not in agreement with the data. This suggests that small corrections should be introduced and the residual symmetries should be slightly broken. 

A common approach to realise the breaking of $G_\text{f}$ is to introduce flavons, new scalars that couple to fermions and have non-trivial properties under the flavour symmetry. These scalars get vacuum expectation values (VEVs), leading to the spontaneous breaking of the flavour symmetry and leaving residual symmetries in the charged lepton and neutrino sectors, respectively. At least two flavon multiplets, one for charged leptons and the other for neutrinos, have to be introduced to guarantee different residual symmetries in the two sectors. The well-known and simplest case is the realisation of TBM in $A_4$ models \cite{Altarelli:2005yp,Altarelli:2005yx,A4_TBM,PZ}. In models with larger symmetry groups, more flavon multiplets may be needed for model constructions \cite{S4_TBM, S4_BM_direct, A5_GR_direct, Delta48, Delta96_TFH_direct}. 

The slight breaking of the residual symmetries can be provided by additional interactions of flavons. The latter may be directly from higher-dimensional operators in the couplings between flavons and leptons \cite{King:2013eh,Altarelli:2010gt}. In our recent paper \cite{PZ}, we observe that cross coupling between neutrino and charged lepton flavons can shift the  VEVs from their original $Z_3$ and $Z_2$ symmetric values. In the models based on $A_4$, we studied in detail the modification to the TBM flavour mixing pattern, in particular conserving the origin of non-zero $\theta_{13}$ and Dirac-type CP violation. 

The interactions of flavons and leptons, in addition to ensuring special Yukawa structures in the lepton sector, may also contribute to other processes and in particular lead to lepton-flavour-violating (LFV) processes. 
Most flavour models assume that the flavour symmetry is broken at a very high scale such that these processes are too suppressed to be observed. However, the scale of flavour symmetry is not known and could be much lower than commonly considered. An electroweak-scale flavour symmetry has recently been discussed \cite{3Higgs_lepton}, see also \cite{3Higgs_quark}. For instance, some flavons are formed by multi-Higgs ($SU(2)_L$ scalar doublets), their VEVs must be below the electroweak scale. If the scale and the flavon masses are sufficiently low, there would be some testable signatures, especially in  charged LFV decay channels. Measuring these processes would provide important clues to identify the origins of leptonic flavour mixing. 

The LFV decays of charged leptons induced by flavons can be divided into two classes: those preserving the residual symmetry in the charged lepton sector and those breaking it. In $A_4$ models, the only processes allowed by the $Z_3$ residual symmetry are $\tau^- \to \mu^+ e^- e^-$, $\tau^- \to e^+ \mu^- \mu^-$, and all other 3-body and radiative decays are forbidden \cite{triality,Muramatsu:2016bda}. The latter can take place if the $Z_3$ symmetry is broken \cite{Kobayashi:2015gwa}, but are typically suppressed due to the consistency with oscillation data, as shown later. 
Current experimental bounds of the branching ratios of LFV $\tau$ 3-body decays $\tau^-\to \mu^+ e^- e^-,\, e^+ \mu^- \mu^-,\,\mu^+\mu^-\mu^-,\,e^+e^-\mu^-,\,\mu^+\mu^-e^-,\,e^+e^-e^-$, and ratiative decays $\tau^-\to \mu^-\gamma,\,e^-\gamma$ are in general around $10^{-8}$, measured by Belle \cite{Belle} and BaBar \cite{BaBar}, respectively. The upper limit of the $\mu$ 3-body decay $\mu^-\to e^+e^-e^-$ decay is $1.0\times10^{-12}$ at 90 \% C.L., from the SINDRUM experiment \cite{SINDRUM}. The most stringent measurement is $\mu^- \to e^- \gamma$ in the MEG experiment, with branching ratio $\sim 4.2 \times 10^{-13}$ at 90\% C.L. \cite{MEG}. A MEG upgrade (MEG II) is envisaged to reach the upper limit of the branching ratio to $4\times 10^{-14}$ in the near future \cite{MEG2}. One may expect these experiments provide important constraints to the scale of flavour symmetry. 

In this paper, we develop a generic method to analyse charged LFV processes in models with discrete flavour symmetries. For definiteness, we choose to work on models based on $A_4$, which we review in section 2. In section 3, we give a model-independent discussion of charged LFV processes induced by flavons. We derive the expressions of leading flavon contributions to $Z_3$-preserving LFV processes and specify different $Z_3$-breaking effects  contributions to $Z_3$-breaking processes. Since the latter are strongly dependent upon the model construction, we list two models and analyse them in detail in section 4. These models, which have been constructed in Ref. \cite{PZ}, are very economical and consistent with current oscillation data.

\section{Flavour mixing in the $A_4$ symmetry}

\subsection{Residual symmetries and tri-bimaximal mixing}

For definiteness, we assume the flavour symmetry is the tetrahedral group $A_4$, which is the group of even permutations of four objects. It is generated by $S$ and $T$ with the requirement $S^2=T^3=(ST)^3=1$, and contains 12 elements: $1$, $S$, $ST$, $TS$, $STS$, $T^2$, $ST^2$, $T^2S$, $TST$, $S$, $T^2ST$, $TST^2$. It is the smallest discrete group which has a 3-dimensional irreducible representation $\mathbf{3}$. In addition, it has three 1-dimensional irreducible representations: $\mathbf{1}$, $\mathbf{1'}$ and $\mathbf{1}''$. The Kronecker product of two 3-dimensional irreducible representations can be reduced as $\mathbf{3}\times\mathbf{3}=\mathbf{1}+\mathbf{1}'+\mathbf{1}''+\mathbf{3}_S+\mathbf{3}_A$, where the subscripts $_S$ and $_A$ stands for the symmetric and anti-symmetric components, respectively. 

We work in the Altarelli-Feruglio basis \cite{Altarelli:2005yx}, where $T$ is diagonal. $T$ and $S$ are respectively given by  
\begin{eqnarray}
T=\left(
\begin{array}{ccc}
 1 & 0 & 0 \\
 0 & \omega ^2 & 0 \\
 0 & 0 & \omega  \\
\end{array}
\right)\,, \qquad
S=\frac{1}{3} \left(
\begin{array}{ccc}
 -1 & 2 & 2 \\
 2 & -1 & 2 \\
 2 & 2 & -1 \\
\end{array}
\right) \,.
\label{eq:generators}
\end{eqnarray}
This basis is widely used in the literature since the charged lepton mass matrix invariant under $T$ is diagonal in this basis. The products of two 3-dimensional irreducible representations $a=(a_1,a_2,a_3)^T$ and $b=(b_1,b_2,b_3)^T$ can be expressed as
\begin{eqnarray}
\hspace{-5mm}
\begin{array}{c}
(ab)_\mathbf{1}\;\, = a_1b_1 + a_2b_3 + a_2b_3 \,,\\
(ab)_\mathbf{1'}\, = a_3b_3 + a_1b_2 + a_2b_1 \,,\\
(ab)_\mathbf{1''} = a_2b_2 + a_1b_3 + a_3b_1 \,,\\
\end{array} \;\;
(ab)_{\mathbf{3}_S} &=& \frac{1}{2} \left(\begin{array}{c} 2a_1b_1-a_2b_3-a_3b_2\\ 2a_3b_3-a_1b_2-a_2b_1\\ 2a_2b_2-a_3b_1-a_1b_3\end{array} \right) , \;\;
(ab)_{\mathbf{3}_A} = \frac{1}{2} \left(\begin{array}{c} a_2b_3-a_3b_2\\ a_1b_2-a_2b_1\\ a_3b_1-a_1b_3 \end{array} \right) .
\label{eq:CG2}
\end{eqnarray}

We assume that $A_4$ is preserved at high energy scale and broken at some lower scale, which we refer to as the scale of flavour symmetry. In the charged lepton and neutrino sectors, residual symmetries $Z_3$ and $Z_2$, which are subgroups of $A_4$, are preserved, respectively. The generators of $Z_3$ and $Z_2$, $T$ and $S$ respectively, generate the full symmetry $A_4$. The invariance of the charged lepton mass matrix under $Z_3$ and that of the neutrino mass matrix under $Z_2$ satisfy
\begin{eqnarray}
T M_l M_l^\dag T^\dag= M_l M_l^\dag\,,\quad
S M_\nu S^T = M_\nu\,.
\end{eqnarray}

%%%%%%%%%%%%%%%%%
\begin{figure}[h!]
\begin{center} 
\includegraphics[width=0.3\textwidth]{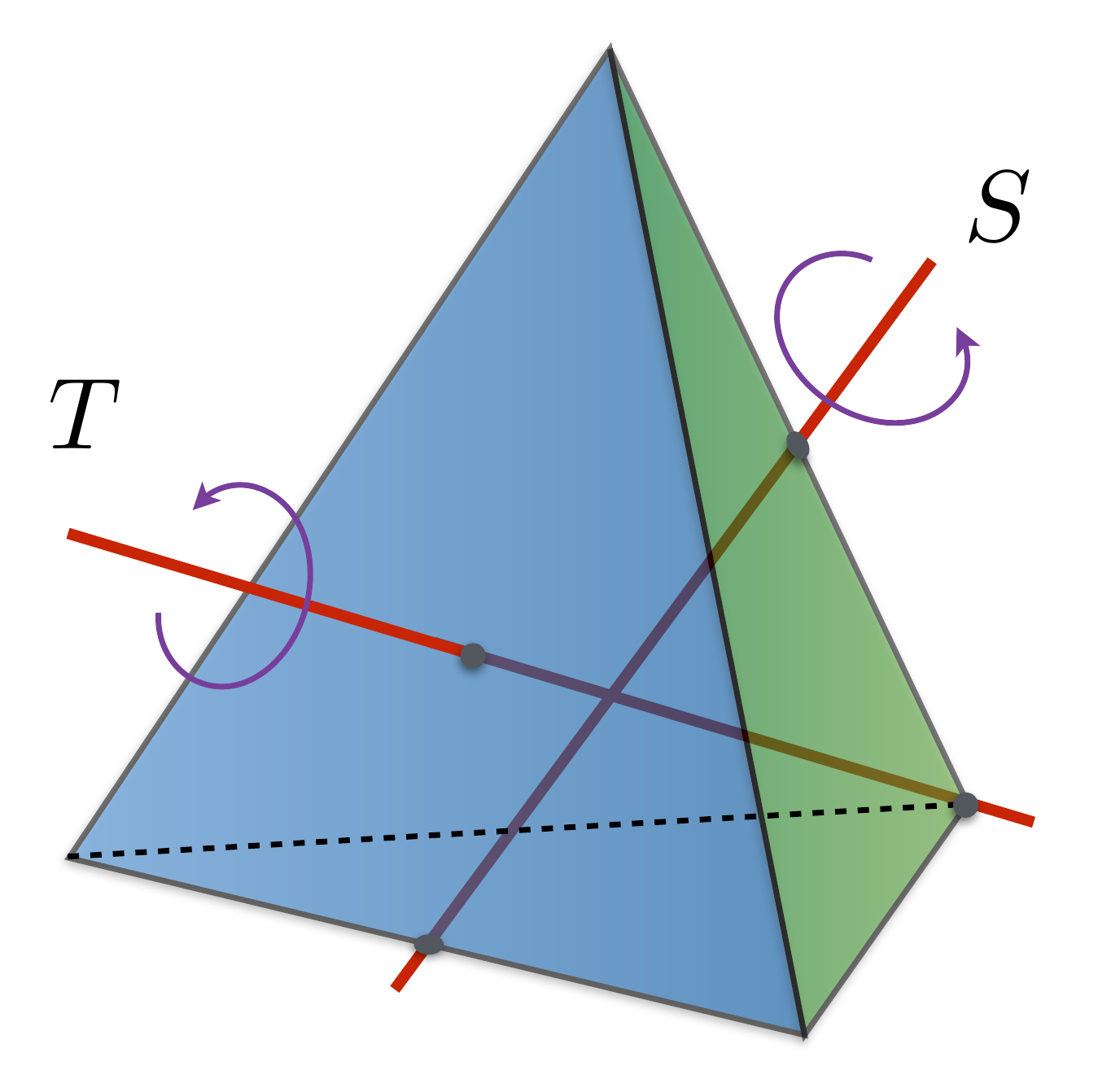}
\vspace{-.5cm} 
\caption{The tetrahedral group $A_4$ as the full flavour symmetry and its subgroups $Z_3=\;<\!\!T\!\!>$ and $Z_2=\;<\!\!S\!\!>$ as residual symmetries in the charged lepton and neutrino sectors, respectively. }
\label{fig:residual}
\end{center} 
\end{figure} 
%%%%%%%%%%%%%%%%%

In order to induce the flavour symmetry breaking, we introduce two flavon triplets $\varphi$, $\chi$ in the charged lepton and neutrino sectors, respectively. We consider two cases with different representation properties:  
\begin{itemize}
\item[i.] The flavons are pseudo-real triplets of $A_4$. In this case, the three components of $\varphi$ satisfy $\varphi_1^*=\varphi_1$ and $\varphi_2^*=\varphi_3$ in the Altarelli-Feruglio basis. This is an economical case introducing as few degrees of freedom beyond the Standard Model as possible. 

\item[ii.] The flavons are complex triplets of $A_4$. This case has been used more widely than case i due to the consistency with supersymmetric and multi-Higgs flavour models and can be regarded as a simplification of these models. 
\end{itemize} 
The representation properties of flavons will have important consequences for LFV processes, as will be discussed later. 

In order to preserve $Z_3$ and $Z_2$, the flavon VEVs should be invariant under the transformations of $T$ and $S$, respectively, i.e.,
\begin{eqnarray}
T \langle \varphi \rangle = \langle \varphi \rangle\,,\qquad
S \langle \chi \rangle = \langle \chi \rangle\,.
\end{eqnarray}
The non-vanishing solutions for the above equation are given by
\begin{eqnarray}
\langle \varphi \rangle = (1,0,0)^T \frac{v_\varphi}{\sqrt{n}}\,,\qquad
\langle \chi \rangle = (1,1,1)^T \frac{v_\chi}{\sqrt{3n}} \,.
\label{eq:vev}
\end{eqnarray}
Here, $n=1,2$ for cases i and ii, respectively, and $v_\varphi$ and $v_\chi$ stand for the overall size of the VEVs and can be treated as $A_4$-breaking scale. Without loss of generality, we assume $v_\varphi,\, v_\chi>0$. 

The Lagrangian terms for generating lepton masses are represented by some higher-dimensional operators. The electroweak lepton doublet $\ell_L=(\ell_{eL},\ell_{\mu L},\ell_{\tau L})^T$ is often arranged to belong to a $\mathbf{3}$ of $A_4$, and the right-handed charged leptons $e_R$, $\mu_R$ and $\tau_R$ belong to singlets $\mathbf{1}$, $\mathbf{1'}$ and $\mathbf{1''}$, respectively.  The relevant Lagrangian terms are given by \footnote{Note that terms such as $(\overline{\ell_L} \chi)_{\mathbf{1}'} \tau_R H$ and $ \big( (\overline{\ell_L} \tilde{H}\tilde{H}^T \ell_L^c)_{\mathbf{3}_S} \varphi  \big)_\mathbf{1}$ cannot be forbidden by $A_4$. These terms modify the mixing structures and should be forbidden at leading order. In concrete models, it can be required by introducing additional discrete Abelian symmetry, which will not be discussed here.} 
\begin{eqnarray}
-\mathcal{L}_l &=& \frac{y_e}{\Lambda} (\overline{\ell_L} \varphi)_\mathbf{1} e_R H + \frac{y_\mu}{\Lambda} (\overline{\ell_L} \varphi)_{\mathbf{1}''} \mu_R H + \frac{y_\tau}{\Lambda} (\overline{\ell_L} \varphi)_{\mathbf{1}'} \tau_R H + \text{h.c.} \,, \nonumber\\
-\mathcal{L}_\nu &=& \frac{ y_1}{2\Lambda\Lambda_\text{W}} \big( (\overline{\ell_L} \tilde{H}\tilde{H}^T \ell_L^c)_{\mathbf{3}_S} \chi  \big)_\mathbf{1}  + \frac{y_2}{2\Lambda_\text{W}} (\overline{\ell_L} \tilde{H}\tilde{H}^T \ell_L^c)_\mathbf{1} + \text{h.c.} \,.
\label{eq:Yukawa_coupling} 
\end{eqnarray}
Here, the Higgs $H$ belongs to $\mathbf{1}$ of $A_4$. $\Lambda$ is a new scale higher than $v_\varphi, v_\chi$. It may be a consequence of the decoupling of some heavy $A_4$ multiplet particles. To generate tiny Majorana neutrino masses, we apply the traditional dimension-5 Weinberg operator $(\overline{\ell_L} \tilde{H}\tilde{H}^T \ell_L^c)$ and $\Lambda_\text{W}$ is the related scale, which may be different from $\Lambda$. After the flavons get the VEVs in Eq.~\eqref{eq:vev}, we obtain the lepton mass matrices
\begin{eqnarray}
M_l=\left(
\begin{array}{ccc}
 y_e & 0 & 0 \\
 0 & y_\mu & 0 \\
 0 & 0 & y_\tau \\
\end{array}
\right)\frac{v v_\varphi}{\sqrt{2n}\Lambda} \,,\qquad
M_\nu=\left(
\begin{array}{ccc}
 2a+b & -a & -a \\
 -a & 2a & -a+b \\
 -a & -a+b & 2a \\
\end{array}
\right) \,,
\label{eq:lepton_mass}
\end{eqnarray}
where $v=246$ GeV is the VEV of the Higgs $H$, $a\equiv y_1v_\chi v^2/(4\sqrt{3n}\Lambda\Lambda_\text{W})$ and $b \equiv y_2 v^2/{2\Lambda_\text{W}}$. At leading order, $M_l$ is diagonal and $M_\nu$ is diagonalised by the unitary matrix
\begin{eqnarray}
U_\text{TBM}=\left(
\begin{array}{ccc}
 \frac{2}{\sqrt{6}} & \frac{1}{\sqrt{3}} & 0 \\
 -\frac{1}{\sqrt{6}} & \frac{1}{\sqrt{3}} & \frac{1}{\sqrt{2}} \\
 -\frac{1}{\sqrt{6}} & \frac{1}{\sqrt{3}} & -\frac{1}{\sqrt{2}} \\
\end{array}
\right)
\end{eqnarray} 
and have eigenvalues $m_1=|3a+b|$, $m_2=|b|$ and $m_3=|3a-b|$. This gives rise to $s_{13}=0$, $s_{12}=1/\sqrt{3}$ and $s_{23}=1/\sqrt{2}$, i.e., the so-called TBM mixing pattern. 

\subsection{The breaking of the residual symmetries}

The TBM mixing has been excluded since it predicts a vanishing $\theta_{13}$. To be consistent with neutrino oscillation data, corrections of order $<0.1$
\begin{eqnarray}
s_{13}=\frac{r}{\sqrt{2}} \,, \qquad s_{12} =\frac{1+s}{\sqrt{3}}\,, \qquad s_{23}= \frac{1+a}{\sqrt{2}}\,,
\end{eqnarray}
must be included. In the Standard Model, TBM is modified by radiative corrections which break the residual symmetries. However, such corrections are too small to produce an $\mathcal{O}(0.1)$ $\theta_{13}$. Couplings with different flavon multiplets provide another origin for flavour mixing corrections. 
In the charged lepton sector, a direct origin of $Z_3$-breaking corrections is the interrupt of the flavon triplet $\chi$. And in the neutrino sector, the $Z_2$-breaking origin is from $\varphi$. 
They may contribute to the Yukawa couplings directly through higher-dimensional operators or indirectly through the shifts of the VEVs induced by cross couplings in the flavon potential. After these corrections are included, the PMNS mixing matrix is parametrised as
\begin{eqnarray}
U_\text{PMNS}=U_l^\dag U_\text{TBM} U_\nu P_\nu\,,
\end{eqnarray}
where $U_l$ and $U_\nu$ are unitary matrices representing corrections in the charged lepton sector and neutrino sector, respectively, and $P_\nu$ is a diagonal phase matrix to render positive neutrino masses.  
In this paper, as we focus on the charged lepton sector, we assume corrections from the neutrino sector to be negligible, i.e., $U_\nu-\mathbf{1} \ll U_l-\mathbf{1}$. Then, the mixing matrix can be simplified to $U_\text{PMNS}=U_l^\dag U_\text{TBM} P_\nu$. 

We consider corrections from higher-dimensional operators in the charged lepton sector, which are written in the following form:
\begin{eqnarray}
-\delta \mathcal{L}_l=\big(\overline{\ell_L} \chi_e \big)_\mathbf{1} e_R H +
\big(\overline{\ell_L} \chi_\mu \big)_{\mathbf{1}''} \mu_R H +
\big(\overline{\ell_L} \chi_\tau \big)_{\mathbf{1}'} \tau_R H + \text{h.c.} \,,
\label{eq:Yukawa_coupling_high}
\end{eqnarray}
where 
\begin{eqnarray}
\chi_e \equiv \sum_{m,n}\frac{y_e^{m,n}}{\Lambda^{m+n}} (\varphi^m \chi^n)_\mathbf{3} \,, \quad
\chi_\mu \equiv \sum_{m,n}\frac{y_\mu^{m,n}}{\Lambda^{m+n}} (\varphi^m \chi^n)_{\mathbf{3}} \,, \quad
\chi_\tau \equiv \sum_{m,n}\frac{y_\tau^{m,n}}{\Lambda^{m+n}} (\varphi^m \chi^n)_{\mathbf{3}}  \,,
\end{eqnarray} 
with $m$, $n$ sum for $m+n \geqslant 2$. $y_{e,\mu,\tau}^{m,n}$ are dimensionless complex coefficients. For $m+n=2$, we obtain the following combinations of VEVs 
\begin{eqnarray}
&&\langle (\varphi\chi)_{\mathbf{3}_S} \rangle \propto  (2,-1,-1)^T \,,\quad
\langle (\varphi\chi)_{\mathbf{3}_A} \rangle \propto  (0,1,-1)^T \,, 
\label{eq:vev2}
\end{eqnarray} 
and $\langle (\chi\chi)_\mathbf{3} \rangle$ vanishes at this order. For $m+n=3$, we get another direction of VEV combinations 
\begin{eqnarray}
\langle (\varphi\varphi)_\mathbf{1}\chi \rangle \propto \langle (\chi\chi)_\mathbf{1}\chi \rangle \propto (1,1,1)^T \,.
\label{eq:vev3}
\end{eqnarray} 
One can prove that any other $Z_3$-breaking combinations of the flavon VEVs must have the directions belonging to one of the aboves. After including these corrections, we obtain the most general form 
\begin{eqnarray} 
\langle \chi_e \rangle \propto(1, \epsilon_{e2}, \epsilon_{e3})\,,\qquad
\langle \chi_\mu \rangle \propto(1, \epsilon_{\mu2}, \epsilon_{\mu3})\,,\qquad
\langle \chi_\tau \rangle \propto(1, \epsilon_{\tau2}, \epsilon_{\tau3})\,.
\end{eqnarray}

Another type of correction comes from the vacuum shift of $\varphi$ due to the $Z_3$-breaking couplings between $\varphi$ and $\chi$. The most general VEVs of $\varphi$ takes the form
\begin{eqnarray}
\varphi=(1,\epsilon_{\varphi2},\epsilon_{\varphi3})^T v_\varphi\,, 
\label{eq:vev_shift}
\end{eqnarray}
where $\epsilon_{\varphi_2}$ and $\epsilon_{\varphi_3}$ stand for the vacuum shift of $\varphi$. 
To calculate the exact expressions of the shifts, we expand the flavon potential around the $Z_3$-invariant VEV $\langle \varphi \rangle = (1,0,0)^T v_\varphi$ and separate it in the $Z_3$-preserving part $V_0(\varphi)$ and the $Z_3$-breaking part $V_1(\varphi)$. $V_0(\varphi)$ would result from the self couplings of $\varphi$, e.g., $\big((\varphi\varphi)_\mathbf{3}(\varphi\varphi)_\mathbf{3}\big)_\mathbf{1}$, and some trivial cross couplings with the other flavons, e.g., $(\varphi\varphi)_\mathbf{1}(\chi\chi)_\mathbf{1}$, after $A_4$ breaking. The $Z_3$-breaking $V_1(\varphi)$ include the cross couplings with other flavon multiplets whose VEVs do not respect the $Z_3$ symmetry, e.g., $(\varphi\varphi)_{\mathbf{1}''}(\chi\chi)_{\mathbf{1}'}$ with $\langle\chi\rangle$ only preserving $Z_2$. 
In the two cases for $\varphi$ we are considering, we can obtain the expressions of the shifts.
\begin{itemize}

\item
In case i, where $\varphi$ is a pseudo-real triplet, the most general $Z_3$-preserving and $Z_3$-breaking terms that are relevant to the vacuum shift at first order are given by 
\begin{eqnarray}
V_0^{(2)}(\varphi) &=& \frac{1}{2}m_{\varphi_1}^2 \varphi_1^2 + m_{\varphi_2}^2 \varphi_2^* \varphi_2    \,, \nonumber\\
V_1^{(1)}(\varphi) &=& \varepsilon_1 v_\varphi^3 \varphi_2^* + \text{h.c.}\,,
\label{eq:potental_vev_shift}
\end{eqnarray}
respectively, where the real parameters $m_{\varphi_1}$, $m_{\varphi_2}$ are the masses of $\varphi_1$ and $\varphi_2$, respectively and $\varepsilon_1$ is a complex dimensionless parameter. The accidential $Z_2'$ symmetry can be recovered if $\varepsilon_1$ is real. 
$V_0^{(2)}(\varphi)$ is invariant under the transformation $\varphi_2\to \omega^2 \varphi_2$, which is required by the $Z_3$ symmetry. The minimisation of $V_0^{(2)}(\varphi)+ V_1^{(1)}(\varphi)$ leads to $\epsilon_{\varphi2}=\epsilon_{\varphi3}^*=\epsilon_\varphi$ with $\epsilon_\varphi$ defined by $\epsilon_\varphi \equiv -\varepsilon_1 v_\varphi^2/m_{\varphi_2}^2$. 

\item
If $\varphi$ is a complex scalar, i.e., case ii, the relevant terms of $\varphi$ are modified to
\begin{eqnarray}
V_0^{(2)}(\varphi) &=& \frac{1}{2}m_{\varphi_1}^2 h_1^2 + m_{\varphi_2}^2 \varphi_2^* \varphi_2 + m_{\varphi_3}^2 \varphi_3^* \varphi_3 + (m_{\varphi_2\varphi_3}^2 \varphi_2 \varphi_3 + \text{h.c.}) \,,\nonumber\\
V_1^{(1)}(\varphi) &=& \varepsilon_1 v_\varphi^3 \varphi_2^* + \varepsilon_1' v_\varphi^3 \varphi_3^* + \text{h.c.} \,,
\label{eq:potental_vev_shift2}
\end{eqnarray}
where $m_{\varphi_1}$, $m_{\varphi_2}$, $m_{\varphi_3}$ are real and $m_{\varphi_2 \varphi_3}$, $\varepsilon_1$ and $\varepsilon_1'$ are in general complex. The phase of $m_{\varphi_2 \varphi_3}$ is unphysical and can always be rotated away by an overall phase redefinition of the flavon $\varphi$. $h_1$ is the real component of $\varphi_1$, $\varphi_1 \equiv (v_\varphi+h_1+i a_1)/\sqrt{2}$, and the pseudo-real scalar $a_1$ becomes an unphysical massless Goldstone particle after $A_4$ breaking to $Z_3$.  From Eq.~\eqref{eq:potental_vev_shift2}, we derive the vacuum shifts
\begin{eqnarray} 
\epsilon_{\varphi2}=-\frac{\varepsilon_1 m_{\varphi_3}^2-\varepsilon_1' m_{\varphi_2\varphi_3}^2}{m_{\varphi_2}^2 m_{\varphi_3}^2 - m_{\varphi_2\varphi_3}^4} v_\varphi^2\,,\qquad
\epsilon_{\varphi3}=-\frac{\varepsilon_1' m_{\varphi_2}^2-\varepsilon_1 m_{\varphi_2\varphi_3}^2}{m_{\varphi_2}^2 m_{\varphi_3}^2 - m_{\varphi_2\varphi_3}^4} v_\varphi^2\,.
\end{eqnarray}
\end{itemize}

After considering the direct corrections to Yukawa couplings from higher-dimensional operators and  the indirect corrections from the flavon vacuum shift, the charged lepton mass matrix becomes non-diagonal: 
\begin{eqnarray} 
M_l=\left(
\begin{array}{ccc}
 y_e & y_\mu (\epsilon_{\mu3}+\epsilon_{\varphi3}) & y_\tau (\epsilon_{\tau2}+\epsilon_{\varphi2}) \\
 y_e (\epsilon_{e2}+\epsilon_{\varphi2}) & y_\mu & y_\tau (\epsilon_{\tau3}+\epsilon_{\varphi3}) \\
 y_e (\epsilon_{e3}+\epsilon_{\varphi3}) & y_\mu (\epsilon_{\mu2}+\epsilon_{\varphi2}) & y_\tau \\
\end{array}
\right)\frac{v v_\varphi}{\sqrt{2}\Lambda} \,,
\label{eq:mass_shift1}
\end{eqnarray} 
leading to the mixing matrix 
\begin{eqnarray}
U_l^\dag=\left(
\begin{array}{ccc}
 1 & -(\epsilon_{\mu3}+\epsilon_{\varphi3}) & -(\epsilon_{\tau2}+\epsilon_{\varphi2}) \\
 \epsilon_{\mu3}^*+\epsilon_{\varphi3}^* & 1 & -(\epsilon_{\tau3}+\epsilon_{\varphi3}) \\
 \epsilon_{\tau2}^*+\epsilon_{\varphi2}^* & \epsilon_{\tau3}^*+\epsilon_{\varphi3}^* & 1 \\
\end{array}
\right) \,.
\label{eq:mixing_shift1} 
\end{eqnarray}
As $e_L$, $\mu_L$ and $\tau_L$ take different $Z_3$ charges, their mixing leads to the breaking of the $Z_3$ symmetry.  
Neglecting the corrections of $U_\nu$, this can be recast in terms of the mixing angles given by
\begin{eqnarray}
&&\sin\theta_{13}=\frac{1}{\sqrt{2}}|\epsilon_{\tau2}-\epsilon_{\mu3}+\epsilon_{\varphi2}-\epsilon_{\varphi3}| \,, \nonumber\\
&&\sin\theta_{12}=\frac{1}{\sqrt{3}}\big[1-\text{Re}(\epsilon_{\tau2}+\epsilon_{\mu3}+\epsilon_{\varphi2}+\epsilon_{\varphi3})\big] \,, \nonumber\\
&&\sin\theta_{23}=\frac{1}{\sqrt{2}}\big[1+\text{Re}(\epsilon_{\tau3}+\epsilon_{\varphi3})\big] \,.  
\end{eqnarray}
The Dirac phase at leading order is given by 
\begin{eqnarray} 
\delta=-\text{Arg}\big\{\epsilon_{\tau2}-\epsilon_{\mu3}+\epsilon_{\varphi2}-\epsilon_{\varphi3} \big\} \,,
\end{eqnarray}
and the Majorana phases cannot be determined. In a specific model, these corrections may not be independent with each other due to additional assumptions, and sum rules of mixing angles and the Dirac phase could appear. 

\section{Charged lepton flavour violation in flavour models}

We now focus on the analysis of LFV decays of charged leptons mediated by flavons, including 3-body decays $l_1^-\to l_2^+ l_3^- l_4^-$ and radiative decays $l_1^-\to l_2^- \gamma$. 
All charged LFV processes mediated by flavons in $A_4$ flavour models can be divided into two parts: those consistent with the $Z_3$ residual symmetry and those violating it. 
The only allowed $Z_3$-preserving LFV decays are $\tau^-\to \mu^+ e^- e^-$ and $\tau^-\to e^+ \mu^- \mu^-$. 
We only focus on charged LFV processes induced by flavons. Namely, these LFV processes originate from the couplings between flavons and charged leptons that generate charged lepton masses and give rise to special flavour structures \footnote{In the most general case, dimension-6 operators 
\begin{eqnarray} 
&&\frac{C_1}{\Lambda^{\prime2}} (\overline{\ell_L} \gamma_\mu \ell_L)_{\mathbf{1}'} (\overline{\ell_L} \gamma^\mu \ell_L)_{\mathbf{1}''}  \,,\quad 
\frac{C_2}{\Lambda^{\prime2}} \big((\overline{\ell_L} \gamma_\mu \ell_L)_{\mathbf{3}} (\overline{\ell_L} \gamma^\mu \ell_L)_{\mathbf{3}} \big)_\mathbf{1}\,, \quad
\frac{C_3}{\Lambda^{\prime2}} (\overline{\ell_L} \gamma_\mu \ell_L)_{\mathbf{1}''} (\overline{e_R} \gamma^\mu \mu_R)\,,  \nonumber\\
&&\frac{C_4}{\Lambda^{\prime2}} (\overline{\ell_L} \gamma_\mu \ell_L)_{\mathbf{1}'} (\overline{e_R} \gamma^\mu \tau_R)\,, \qquad
\frac{C_5}{\Lambda^{\prime2}} (\overline{e_R} \gamma_\mu \mu_R) (\overline{e_R} \gamma^\mu \tau_R)\,, \qquad\quad
\frac{C_6}{\Lambda^{\prime2}} (\overline{\mu_R} \gamma_\mu e_R) (\overline{\mu_R} \gamma^\mu \tau_R)\,.
\end{eqnarray} 
cannot be forbidden by $A_4$ and the electroweak symmetry, and allow the $Z_3$-preserving LFV decays. 
Assuming $C_i\sim \mathcal{O}(1)$, the experimental constraint to the scale $\Lambda^{\prime}$ is $\Lambda'>15$ TeV \cite{Feruglio:2008ht}. However, they are not essential to generate lepton mass matrices with flavour structures, and thus will not be discussed in this paper. }.

Flavon fields couple to charged leptons, as shown in Eq.~\eqref{eq:Yukawa_coupling} with subleading-order corrections shown in Eq.~\eqref{eq:Yukawa_coupling_high}. These couplings are suppressed by charged lepton masses. After the flavour symmetry breaking, the flavon fields gain VEVs, masses and mixing, from the potential $V(\varphi)$. 
Generically, one can write out the effective operators of the 3-body LFV decay $l_1^-\to l_2^+ l_3^- l_4^-$ after the breakings of the flavour symmetry and the electroweak symmetry as 
\begin{eqnarray}
\mathcal{L}^{(6)} &=& \sum_{P_i}
C^{l_4 l_2 l_3 l_1}_{P_4P_2P_3P_1} (\overline{l_{4P_4}}l_{2P_2})(\overline{l_{3P_3}}l_{1P_1}) + C^{l_3 l_2 l_4 l_1}_{P_3P_2P_4P_1} (\overline{l_{3P_3}}l_{2P_2})(\overline{l_{4P_4}}l_{1P_1}) \,,
\label{eq:d6_operators}
\end{eqnarray}  
where $P_i=L,R$ (for $i=1,2,3,4$) and the coefficients $C^{l_4 l_2 l_3 l_1}_{P_4P_2P_3P_1}$, $C^{l_3 l_2 l_4 l_1}_{P_3P_2P_4P_1}$ are functions of charged lepton and flavon mass parameters. 
Later we will see that due to choices of representations $e_R\sim\mathbf{1}$, $\mu_R\sim\mathbf{1'}$, $\tau_R\sim\mathbf{1''}$ and the large hierarchy $m_e \ll m_\mu \ll m_\tau$, the contribution corresponding to $P_1=L$ is subleading in the $A_4$ models and can be neglected in our discussion. Ignoring charged lepton masses in the final states, we derive the decay width of $l_1^-\to l_2^+ l_3^- l_4^-$ as 
\begin{eqnarray} 
\Gamma(l_1^-\to l_2^+ l_3^- l_4^-) &\approx& \frac{\eta m_{l_1}^5}{3(16\pi)^3} \Big[ \big|C^{l_4 l_2 l_3 l_1}_{LRLR}\big|^2 + \big|C^{l_3 l_2 l_4 l_1}_{LRLR} \big|^2 - \text{Re}\Big(C^{l_4 l_2 l_3 l_1}_{LRLR} (C^{l_3 l_2 l_4 l_1}_{LRLR})^*\Big)  \nonumber\\
&& \hspace{12.5mm}+ \big|C^{l_4 l_2 l_3 l_1}_{RLLR}\big|^2 + \big|C^{l_3 l_2 l_4 l_1}_{RLLR} \big|^2 - \text{Re}\Big(C^{l_4 l_2 l_3 l_1}_{RLLR} (C^{l_3 l_2 l_4 l_1}_{RLLR})^*\Big) \Big] \,, 
\end{eqnarray} 
where $\eta=1,2$ for $l_3 = l_4$, $l_3\neq l_4$, respectively and $m_{l_1}$ is the mass of $l_1$. 
As for the radiative decay $l_1^-\to l_2^-\gamma$, its amplitude is generically written as $ \overline{u_{l_2}} \Gamma_{\mu}^{l_2l_1} u_{l_1} \epsilon^{\mu*} $ \cite{radiative} with
\begin{eqnarray}
\Gamma_{\mu}^{l_2l_1} = i \sigma_{\mu\nu} q^\nu (A_L^{l_2l_1} P_L+ A_R^{l_2l_1} P_R) \,,
\label{eq:EDM}
\end{eqnarray}
where $R_{L,R}=(1\mp\gamma_5)/2$, and the coefficients $A_L^{l_2l_1}$,~$A_R^{l_2l_1}$ are dependent upon charged lepton and flavon mass parameters. We mention that also due to choices of representations of $e_R$, $\mu_R$, $\tau_R$ and the large hierarchy of charged lepton masses, $A_L^{l_2l_1} \ll A_R^{l_2l_1}$. Thus, the decay rate can be expressed as
\begin{eqnarray}
\Gamma(l_1^-\to l_2^-\gamma) = \frac{m_{l_1}^3}{16\pi} |A_R^{l_2l_1}|^2\,.
\end{eqnarray}

\subsection{$Z_3$-preserving LFV charged lepton decays} 

From the Lagrangian terms in Eq.~\eqref{eq:Yukawa_coupling}, we can write the couplings between flavon and charged leptons explicitly. In the Altarelli-Feruglio basis, they are given by
\begin{eqnarray}
\mathcal{L}_l^\text{eff}&=&\frac{m_e}{v_\varphi} \left( \,\overline{e_L} e_R\, \varphi_1 + \overline{\mu_L} e_R \varphi_2 +\overline{\tau_L} e_R \varphi_3 \right) \sqrt{n} \nonumber\\
&+&\frac{m_\mu}{v_\varphi} \left( \overline{\mu_L} \mu_R \varphi_1 + \overline{\tau_L} \mu_R \varphi_2 +\overline{e_L} \mu_R \varphi_3 \right) \sqrt{n} \nonumber\\
&+&\frac{m_\tau}{v_\varphi} \left( \,\overline{\tau_L} \tau_R \, \varphi_1 + \, \overline{e_L} \tau_R \varphi_2 +\overline{\mu_L} \tau_R \varphi_3 \right) \sqrt{n} + \text{h.c.}\,.
\label{eq:Feynman}
\end{eqnarray}
The $Z_3$ symmetry corresponds to the invariance under the transformation 
\begin{eqnarray} 
\hspace{-3mm}(e_L,\,e_R, \,\varphi_1) \to (e_L,\,e_R, \,\varphi_1) \,,\quad
(\mu_L,\,\mu_R, \,\varphi_2) \to \omega^2 (\mu_L,\,\mu_R, \,\varphi_2)\,,\quad
(\tau_L,\, \tau_R, \,\varphi_3) \to \omega (\tau_L,\, \tau_R,\, \varphi_3) \,.
\end{eqnarray} 
Namely, $e_L$, $e_R$, $\varphi_1$ are invariant under $Z_3$, $\tau_L$, $\tau_R$, $\varphi_3$ are covariant under $Z_3$ with a charge 1, and $\mu_L$, $\mu_R$, $\varphi_2$ are covariant under $Z_3$ with a charge 2 (or equivalently, contravariant under $Z_3$ with a charge 1). 
In the case of transfer momentum much lower than the scale of flavour symmetry and flavon masses, one can integrate out $\varphi_{1}$, $\varphi_2$, $\varphi_3$, and derive the effective 4-fermion interactions. While the $Z_3$-invariant flavon $\varphi_1$ induces flavour-conserving processes, the $Z_3$-covariant flavons $\varphi_2$ and $\varphi_3$ are the main sources for charged LFV processes. 
As $e$, $\mu$ and $\tau$ take different $Z_3$ charges in $A_4$ models, it is easy to prove that the only allowed processes are $\tau^-\to \mu^+ e^-e^-$ and $\tau^-\to e^+ \mu^- \mu^-$. The other 3-body decay and all radiative decay modes are forbidden at this level \cite{triality}. 

In case i, we recall that $\varphi$ is a pseudo-real triplet, $\varphi_1^*=\varphi_1$, $\varphi_2^*=\varphi_3$. As shown in Eq.~\eqref{eq:potental_vev_shift}, $\varphi_1$ and $\varphi_2$ have different masses $m_{\varphi_1}^2$ and $m_{\varphi_2}^2$, and the $Z_3$ symmetry forbids the mixing between them at leading order. 
The 4-fermion interactions mediated by $\varphi_1$ and $\varphi_2$ are given by
\begin{eqnarray}
\mathcal{L}_{\varphi_1}&=& 
\frac{1}{m_{\varphi_1}^2}\Big[ 
\frac{m_e}{v_\varphi}\overline{e}e+ 
\frac{m_\mu}{v_\varphi}\overline{\mu}\mu+ 
\frac{m_\tau}{v_\varphi}\overline{\tau}\tau\Big]^2 \,, \nonumber\\
\mathcal{L}_{\varphi_2}&=&
\frac{1}{m_{\varphi_2}^2}\Big[
\frac{m_e}{v_\varphi}(\overline{\mu_L}e_R+\overline{e_R}\tau_L)+ 
\frac{m_\mu}{v_\varphi}(\overline{\tau_L}\mu_R+\overline{\mu_R}e_L)+
\frac{m_\tau}{v_\varphi}(\overline{e_L}\tau_R+\overline{\tau_R}\mu_L)\Big] \nonumber\\
&&\hspace{5mm}\times\Big[
\frac{m_e}{v_\varphi}(\overline{e_R}\mu_L+\overline{\tau_L}e_R)+ 
\frac{m_\mu}{v_\varphi}(\overline{\mu_R}\tau_L+\overline{e_L}\mu_R)+
\frac{m_\tau}{v_\varphi}(\overline{\tau_R}e_L+\overline{\mu_L}\tau_R)\Big] \,.
\end{eqnarray}
The above Lagrangian terms are compatible with the flavour triality \cite{triality}, which gives rise to the LFV decay modes $\tau^-\to \mu^+ e^-e^- $ and $\tau^- \to e^+ \mu^-\mu^-$ and their charged-conjugate processes. These are the only 3-body LFV charged lepton decays allowed by the $Z_3$ symmetry. In Model I, operators for $\tau^-\to \mu^+ e^-e^- $ and $\tau^- \to e^+ \mu^-\mu^-$ can be expressed as 
\begin{eqnarray}
\frac{m_\mu m_\tau}{v_\varphi^2m_{\varphi_2}^2} (\overline{e_L}\mu_R)(\overline{e_L}\tau_R) \,,\quad
\frac{m_\mu m_\tau}{v_\varphi^2m_{\varphi_2}^2} (\overline{\mu_R}e_L)(\overline{\mu_L}\tau_R) \,,
\label{eq:CLFV_Z3_ModelI}
\end{eqnarray}
respectively, with both coefficients $C^{e\mu e\tau}_{LRLR}$ and $C^{\mu e\mu\tau}_{RLLR}$ suppressed by $\mu$ and $\tau$ masses.
We just list the leading contribution here. Terms such as $(\overline{e_R}\mu_L)(\overline{e_L}\tau_R)$ and $(\overline{\mu_R}e_L)(\overline{\mu_R}\tau_L)$ are also allowed, but sub-leading, suppressed by $\frac{m_e m_\tau}{v_\varphi^2m_{\varphi_2}^2}$ or $\frac{m_\mu^2}{v_\varphi^2m_{\varphi_2}^2}$, and will not be considered in the following. Then, we get approximatively equal branching ratios of these two processes
\begin{eqnarray}
\text{Br}(\tau^-\to \mu^+ e^-e^-) \approx \text{Br}(\tau^-\to e^+ \mu^- \mu^-) \,, 
\label{eq:branching_sum_rule1}
\end{eqnarray}
both suppressed by $\big(\frac{m_\mu m_\tau v^2}{m_{\varphi_2}^2v_\varphi^2}\big)^2$. 
If we assume the scale of flavour symmetry $v_\varphi$ and flavon masses to be around the electroweak scale $v=246$ GeV, the branching ratios will be smaller than $10^{-11}$. 
One highlighted feature is that both $Z_3$-preserving processes have the same branching ratios. 
This is because $\overline{e_L} \tau_R \varphi_2$ and $\overline{\mu_L} \tau_R \varphi_3$ have the same coefficient as shown in Eq. \eqref{eq:Feynman}, and $\varphi_2$ is the complex conjugate of $\varphi_3$. 
Essentially, it is a consequence of the $Z_3$ symmetry remaining from the breaking of $A_4$ and the economical choice of the pseudo-real presentation of $\varphi$. 

In case ii, as $\varphi$ is a complex triplet of $A_4$, the mixing between the $Z_3$-covariant flavons $\varphi_2$ and $\varphi_3^*$ should be considered in the $Z_3$-preserving processes. As shown in Eq.~\eqref{eq:potental_vev_shift2}, the $Z_3$ symmetry cannot forbid the off-diagonal mass term $m_{\varphi_2\varphi_3}^2 (\varphi_2 \varphi_3 + \text{h.c.})$. We need to go into the mass basis $\varphi_2'$ and $\varphi_3'$ with mass eigenvalues $m_{\varphi_2'}$ and $m_{\varphi_3'}$, thanks to the rotation: 
\begin{eqnarray}
\left(\begin{array}{c} \varphi_2^{\prime} \\ \varphi_3^{\prime*} \end{array} \right)  = 
\left(\begin{array}{cc}
 c_\vartheta & -s_\vartheta \\
 s_\vartheta & c_\vartheta \\
\end{array}\right)
\left(\begin{array}{c} \varphi_2 \\ \varphi_3^{*} \end{array} \right) \,,
\end{eqnarray}
with $s_\vartheta\equiv\sin\vartheta$, $c_\vartheta\equiv\cos\vartheta$ and $\tan2\vartheta=2m_{\varphi_2\varphi_3}^2/(m_{\varphi_3}^2-m_{\varphi_2}^2)$, $-45^\circ<\vartheta\leqslant45^\circ$. Then, integrating the massive scalars out, we obtain the coefficients of the 4-fermion processes $\tau^-\to \mu^+ e^-e^-$ and $\tau^- \to e^+ \mu^-\mu^-$ as
\begin{eqnarray}
C^{e\mu e\tau}_{LRLR} =
 \frac{m_\mu m_\tau}{v_\varphi^2} \left(\frac{\sin2\vartheta}{m_{\varphi_3^{\prime}}^2}-\frac{\sin2\vartheta}{m_{\varphi_2^{\prime}}^2}\right) \,, \qquad
C^{\mu e\mu\tau}_{RLLR} = 
\frac{m_\mu m_\tau}{v_\varphi^2}\left(\frac{2s^2_\vartheta}{m_{\varphi_2^{\prime}}^2}+\frac{2c^2_\vartheta}{m_{\varphi_3^{\prime}}^2}\right)  \,, 
\end{eqnarray}
respectively. 
Compared with case i, where the coefficients of $(\overline{e_L}\mu_R)(\overline{e_L}\tau_R)$ and $(\overline{\mu_R}e_L)(\overline{\mu_L}\tau_R)$ are the same, the coefficients in case ii are in general different. This feature could be used to establish if $\varphi$ is in a pseudo-real or complex representation if future experiments observe the signatures. 
We emphasise that since the mixing between $\varphi_2$ and $\varphi_3^{*}$ is in general large, even maximal, branching ratios for both channels should be at the same level, both proportional to $m_\mu m_\tau$. 
The above equation reduces to Eq.~\eqref{eq:CLFV_Z3_ModelI} in the limit $m_{\varphi_2^{\prime}}^2\to \infty$ and $\vartheta\to 45^\circ$, or $m_{\varphi_3^{\prime}}^2\to \infty$ and $\vartheta\to -45^\circ$. 
Only in the limit $\vartheta \to 0$, $\tau^-\to \mu^+  e^-e^-$ is suppressed, as discussed in \cite{Kobayashi:2015gwa}.

\subsection{$Z_3$-breaking LFV charged lepton decays}

The $Z_3$-breaking LFV processes have three sources, depending on their connection with flavour mixing. 
One is the mixing of charged lepton mass eigenstates, characterised by $\epsilon_{\mu3}+\epsilon_{\varphi3}$, $\epsilon_{\tau2}+\epsilon_{\varphi2}$ and $\epsilon_{\tau3}+\epsilon_{\varphi3}$ in the last section. The left-handed charged lepton mass eigenstates are superpositions of $e_L$, $\mu_L$ and $\tau_L$, which obviously break the $Z_3$ symmetry: 
\begin{eqnarray}
\left( \begin{array}{c}
 e_L \\ \mu_L \\ \tau_L \\
\end{array}\right) 
\to U_l 
\left( \begin{array}{c}
 e_L \\ \mu_L \\ \tau_L \\
\end{array}\right) \,.
\label{eq:lepton_shift} 
\end{eqnarray}
There is also mixing of the right-handed charged leptons $e_R$, $\mu_R$ and $\tau_R$, but their mixing is suppressed by both $\epsilon_\varphi$ and the hierarchy of charged lepton masses \footnote{This is due to the typical feature in $A_4$ models that $e_R$, $\mu_R$ and $\tau_R$ are always arranged as singlets $\mathbf{1}$, $\mathbf{1}'$, $\mathbf{1}''$ of $A_4$.}, and thus can be safely neglected. Charged LFV processes induced by this effect is easy to be calculated. 

The other two are related to the $Z_3$-breaking property of the flavon triplet $\varphi$. 
When the vacuum shift results in the mixing between charged leptons, it also results in the mixing and mass corrections of different components of $\varphi$. The mixing between the $Z_3$-invariant flavon $\varphi_1$ and the $Z_3$-covariant flavons $\varphi_2$ and $\varphi_3$, and the mass splitting of the two real degrees of freedom for each $Z_3$-covariant complex flavons contribute to LFV processes.  

To calculate the mass corrections to and the mixing of $\varphi$, we expand the $Z_3$-preserving potential $V_0(\varphi)$ to the third order and the $Z_3$-breaking potential $V_1(\varphi)$ to the second order around the $Z_3$-invariant VEV $\langle \varphi \rangle = (1,0,0)^T v_\varphi$. Taking a pseudo-real flavon triplet as an example, the  $Z_3$-preserving terms are expressed as 
\begin{eqnarray}
V_0^{(3)}(\varphi) &=&  \frac{1}{3} k_1 v_\varphi \varphi_1^3 + k_2 v_\varphi \varphi_1\varphi_2^*\varphi_2  + \frac{1}{3} v_\varphi (k_3\varphi_2^3 + k_3^* \varphi_2^{*3}) \,. 
\label{eq:cubic_V}
\end{eqnarray}
Here, $k_1$, $k_2$ are real, required by the Hermiticity of the potential. If the potential is renormalisable, $k_3$ is also real, since all the renormalisable $A_4$-invariant combinations of $\varphi$, including $(\varphi \varphi)_\mathbf{1}$, $\big((\varphi \varphi)_{\mathbf{3}_S} \varphi \big)_\mathbf{1}$, $(\varphi \varphi)_{\mathbf{1}'} (\varphi \varphi)_{\mathbf{1}''} $, $\big((\varphi \varphi)_{\mathbf{3}_S} (\varphi \varphi)_{\mathbf{3}_S} \big)_\mathbf{1}$, are real. Once higher-dimensional operators are included in the potential, e.g., $\lambda\big((\varphi \varphi)_{\mathbf{1}'}\big)^3+ \lambda^*\big((\varphi \varphi)_{\mathbf{1}''}\big)^3$, $k_3$ can be complex. 
Since the vacuum is shifted, a small mass term which splits the masses of two components of the complex $\varphi_2$ and a mixing term between $\varphi_1$ and $\varphi_2$ can be generated by the cubic couplings proportional to $k_2$ and $k_3$ in Eq.~\eqref{eq:cubic_V}, respectively. After the flavon VEV shifts from $\langle \varphi \rangle = (1,0,0)^T v_\varphi$ to $\langle \varphi \rangle = (1,\epsilon_\varphi,\epsilon_\varphi^*)^T v_\varphi$, the cubic term $V_0^{(3)}(\varphi)$ will contribute a small mass term for the flavons 
\begin{eqnarray}
\delta V_0^{(2)}=\epsilon_\varphi v_\varphi^2 ( k_2 \varphi_1 \varphi_2^* + k_3 \varphi_2^2 ) + \text{h.c.} \,. 
\end{eqnarray}
The quadratic terms of $V_1(\varphi)$ can also generate such mass terms. In general, they are given by 
\begin{eqnarray}
V_1^{(2)}(\varphi) &=& \varepsilon_2 v_\varphi^2 \varphi_1 \varphi_2^* + \varepsilon_3 v_\varphi^2 \varphi_2^2 + \text{h.c.}  \,, 
\label{eq:dV}
\end{eqnarray}
where $\varepsilon_2$ and $\varepsilon_3$ are in general complex parameters. 
$V_1^{(2)}(\varphi)$ obviously can originate from cross couplings between $\varphi$ and $\chi$, including the renormalisable terms, i.e., $(\varphi \chi)_\mathbf{1}$, $\big((\varphi \varphi)_{\mathbf{3}_S} \chi \big)_\mathbf{1}$, $(\varphi \varphi)_{\mathbf{1}''} (\chi \chi)_{\mathbf{1}'} $, and the higher-dimensional terms, i.e., $\big(\big((\varphi \varphi)_{\mathbf{3}_S} (\varphi \varphi)_{\mathbf{3}_S} \big)_\mathbf{\mathbf{3}_S}\varphi\big)_{\mathbf{1}}$, $[(\varphi \varphi)_{\mathbf{1}'}]^2 (\chi \chi)_{\mathbf{1}'}$. The $\varepsilon_2$ and $\varepsilon_3$ terms can give another mass term,  splitting masses of the two components of $\varphi_2$ and mixing of $\varphi_1$ and $\varphi_2$, respectively. 

We include the contributions from $\delta V_0^{(2)}(\varphi)$ and $V_1^{(2)}(\varphi)$, as well as the leading order result $V_0^{(2)}(\varphi)$. The two $Z_3$-breaking effects of $\varphi$ mentioned above can be discussed analytically. 
The first one is the mixing between the $Z_3$-invariant ($\varphi_1$) and $Z_3$-covariant flavons  ($\varphi_2$). To rotate it to the mass eigenstates of flavons, we need do the following transformation: 
\begin{eqnarray} 
&&\varphi_1\to \varphi_1+ \left( \epsilon_{\varphi_1 \varphi_2}^* \varphi_2 + \epsilon_{\varphi_1 \varphi_2} \varphi_2^*\right)  \,, \nonumber\\ 
&&\varphi_2\to \varphi_2- \epsilon_{\varphi_1 \varphi_2}  \varphi_1 \,,
\label{eq:flavon_shift}
\end{eqnarray}
where 
\begin{eqnarray}
\epsilon_{\varphi_1 \varphi_2}=\frac{( k_2 \epsilon_\varphi + \varepsilon_2 ) v_\varphi^2}{m_{\varphi_2}^2-m_{\varphi_1}^2} \,.
\end{eqnarray}
The second one is the mass splitting of the two components of the complex scalar $\varphi_2$: 
\begin{eqnarray}
m^2_{h_2,a_2}=m_{\varphi_2}^2(1\pm 2 |\epsilon_{h_2 a_2}|) \,.
\label{eq:real_masses}
\end{eqnarray}
with the two components $h_2$ and $a_2$ defined by 
\begin{eqnarray}
h_2=\frac{1}{\sqrt{2}} \Big(\varphi_2 \exp(i\frac{\theta_{h_2 a_2}}{2}) + \varphi_2^* \exp(-i\frac{\theta_{h_2 a_2}}{2})\Big) \,,\nonumber\\
a_2=\frac{-i}{\sqrt{2}} \Big(\varphi_2 \exp(i\frac{\theta_{h_2 a_2}}{2}) - \varphi_2^* \exp(-i\frac{\theta_{h_2 a_2}}{2})\Big) \,,
\label{eq:varphi_real}
\end{eqnarray}
$\epsilon_{h_2 a_2}=k_3 \epsilon_\varphi+ \varepsilon_3$ and $\theta_{h_2 a_2}$ being the phase of $\epsilon_{h_2 a_2}$. 

Branching ratios for the $Z_3$-breaking charged LFV processes can be derived after including the three $Z_3$-breaking sources: the correction to the mixing of charged leptons, the correction to the mixing of flavons and flavon mass splitting. 
For the pseudo-real triplet flavon in case i, these three effects are analytically listed in Eqs.~\eqref{eq:lepton_shift}, \eqref{eq:flavon_shift} and \eqref{eq:real_masses}.
For the complex triplet flavon in case ii, the flavon potential is more complicated and it is hard to extract the general analytical expressions for the later two $Z_3$-breaking sources. 
By imposing additional assumptions, e.g., some Abelian symmetries, can simplify the flavon potential and also the $Z_3$-breaking sources. This scenario will be discussed in the second model in the next section.

\section{$Z_3$-breaking LFV charged lepton decays in concrete models \label{sec:CLFV_models}}

In this section, we will study $Z_3$-breaking LFV charged lepton decays in two models. These models have been proposed and the phenomenology of their flavour mixing has been studied in detail in Ref. \cite{PZ}. After a brief introduction of these models, we will calculate contributions of flavons to the $Z_3$-breaking processes and discuss the experimental constraints. For more details about the model constructions and properties of flavour mixings in both models, we refer to \cite{PZ}. 

\subsection{Model constructions}

The flavour symmetry is assumed to be $A_4\times Z_2^\varphi\times Z_4^\chi$ in both models, with field contents listed in Table \ref{tab:fields}. One right-handed neutrino triplet $N$ and two flavon triplets $\varphi, \chi$ and one flavon singlet $\eta$ of $A_4$ are introduced. The singlet $\eta$ is used to obtain the correct neutrino mass spectrum. The only difference between these two models is that the flavon multiplets in Model I are pseudo-real representations of $A_4$ (case i), while all flavon multiplets in Model II are in complex representations (case ii). 

%%%%%%%%%%%%%%%%%%%%

\begin{table}[h] 
\begin{center} 
\begin{tabular}{c c c c c c c c c c}
\hline\hline
Fields & $\ell_L$ & $e_R,\mu_R,\tau_R$ & $N$ & $H$ & $\varphi$
& $\chi$ & $\eta$    \\ \hline

$A_4$ & $\mathbf{3}$ & $\mathbf{1},\mathbf{1}',\mathbf{1}''$ & $\mathbf{3}$ & $\mathbf{1}$ & $\mathbf{3}$ & $\mathbf{3}$ & $\mathbf{1}$  \\

$Z_2^\varphi$ & $1$ & $-1$ & $1$ & $1$ & $-1$ & $1$ & $1$  \\

$Z_4^\chi$ & $i$ & $i$ & $i$ & $1$ & $1$ & $-1$ & $-1$  \\
\hline 
\hline 
\end{tabular}
\caption{\label{tab:fields} Transformation properties of fields in the flavour symmetry $A_4 \times Z_2^\varphi \times Z_4^\chi$.}
\end{center}
\end{table}

\subsubsection*{Model I}

The most general renormalisable flavon potential of $\varphi$ invariant under the symmetry is written as
\begin{eqnarray}
V(\varphi)= \frac{1}{2}\mu^2_\varphi (\varphi \varphi)_\mathbf{1} +\frac{1}{4} \left[f_1 \big( (\varphi \varphi)_\mathbf{1}\big)^2 + f_2 (\varphi \varphi)_{\mathbf{1}'} (\varphi \varphi)_{\mathbf{1}''} + f_3 \big( (\varphi \varphi)_{\mathbf{3}_S} (\varphi \varphi)_{\mathbf{3}_S} \big)_\mathbf{1} \right]\,,
\label{eq:Vvarphi}
\end{eqnarray}
in which all the coefficients $\mu_\varphi^2$ and $f_{1,2,3}$ are real. 
Once the relations $\mu_\varphi^2<0$ and $f_2>f_3>-f_1$ are required, we can derive the VEV $\langle \varphi \rangle$ in Eq.~\eqref{eq:vev}. The masses of $\varphi_1$ and $\varphi_2$ are given by
\begin{eqnarray}
m_{\varphi_1}^2 = 2(f_1+f_3)v_\varphi^2 \,, &&
m_{\varphi_2}^2 = (f_2-f_3) v_\varphi^2\,. 
\label{eq:flavon_mass1}
\end{eqnarray} 
With replacements $\varphi \to \chi$ and $f_i \to g_i$ in Eq.~\eqref{eq:Vvarphi}, we get the potential of $\chi$. By assuming a different relation $g_3>g_2>-g_1$, we obtain the VEV $\langle \chi \rangle$ in Eq.~\eqref{eq:vev}. All the cross couplings between $\varphi$ and $\chi$ are expressed as 
\begin{eqnarray} 
V(\varphi, \chi) &=& \frac{1}{2} \epsilon_1( \varphi \varphi )_\mathbf{1} ( \chi \chi )_\mathbf{1} + \frac{1}{4}\big[ \epsilon_2 ( \varphi \varphi )_{\mathbf{1}''} ( \chi \chi )_{\mathbf{1}'} +\text{h.c.} \big]
+ \frac{1}{2} \epsilon_3 \big( ( \varphi \varphi )_{\mathbf{3}_S} (\chi \chi )_{\mathbf{3}_S} \big)_{\mathbf{1}} \,,
\label{eq:Vmix}
\end{eqnarray}
where $\epsilon_1$ and $\epsilon_3$ are real and $\epsilon_2$ is complex, and the $\epsilon_2$ term is the only term that will break the $Z_3$ residual symmetry at first order. The relation of $\epsilon_2$ and $\varepsilon_1$ in Eq.~\eqref{eq:potental_vev_shift} is given by $\varepsilon_1=\frac{1}{2}\epsilon_2 v_\chi^2/v_\varphi^2$. The $\epsilon_3$ term will contribute to the breaking of $Z_2$ in $\langle \chi \rangle$. By assuming the VEV $v_\chi$ to be significantly higher than $v_\varphi$, e.g., $|v_\chi| \gtrsim 2 |v_\varphi|$, this contribution can be neglected. 

The Lagrangian generating lepton masses is the same as in Eq.~\eqref{eq:Yukawa_coupling}. Here, we assume the contribution from higher-dimensional operators to be negligible and all the correction to TBM come from one single complex parameter $\epsilon_\varphi$. In this case, the mixing parameters are simplified to \cite{PZ} 
\begin{eqnarray} 
&&\sin\theta_{13}=\sqrt{2}|\epsilon_\varphi\sin\theta_{\varphi}|\,,\nonumber\\
&&\sin\theta_{12}= \frac{1}{\sqrt{3}} \big( 1-2|\epsilon_\varphi|\cos\theta_\varphi \big) \,,\nonumber\\
&&\sin\theta_{23}=\frac{1}{\sqrt{2}} \big( 1+|\epsilon_\varphi|\cos\theta_\varphi \big) \,,\nonumber\\
&&\delta=\left\{\begin{array}{c} 270^\circ -2 |\epsilon_\varphi| \sin\theta_\varphi\,, \quad \theta_\varphi>0 \,, \\ \;\;90^\circ -2 |\epsilon_\varphi| \sin\theta_\varphi\,, \quad \theta_\varphi<0 \,.
 \end{array} \right. 
\label{eq:mixing_Model_I}
\end{eqnarray} 
From the above expression, we see that both $\theta_{13}$ and $\delta$ originate from the same source, the imaginary part of $\epsilon_2$, and almost-maximal CP violation is predicted, with $\delta+\sqrt{2}\theta_{13}\approx 90^\circ,\,270^\circ$. 
In addition, there are sum rules of mixing angles 
\begin{eqnarray}
r^2+s^2=4|\epsilon_\varphi|^2 \,,\quad s+2a=0 \,.
\end{eqnarray}
As shown in Ref. \cite{PZ}, this scenario is compatible with current neutrino oscillation data in the case $r\gg s$, which results in $|\epsilon_\varphi| \approx r/2 \approx 0.1$, or equivalently,  $|\epsilon_\varphi| \approx \theta_{13}/\sqrt{2}$. 

\subsubsection*{Model II } 

In this model, the flavon multiplets are in complex representations. The potential for $\varphi$ is altered to 
\begin{eqnarray}
V(\varphi)&=&\mu^2_\varphi (\tilde{\varphi} \varphi)_\mathbf{1} + f_1 \big( (\tilde{\varphi} \varphi)_\mathbf{1}\big)^2 + f_2 (\tilde{\varphi} \varphi)_{\mathbf{1}'} (\tilde{\varphi} \varphi)_{\mathbf{1}''} + f_3 \big( (\tilde{\varphi} \varphi)_{\mathbf{3}_S} (\tilde{\varphi} \varphi)_{\mathbf{3}_S} \big)_\mathbf{1}  \nonumber\\
&& + f_4 \big( (\tilde{\varphi} \varphi)_{\mathbf{3}_A} (\tilde{\varphi} \varphi)_{\mathbf{3}_A} \big)_\mathbf{1} + f_5 \big( (\tilde{\varphi} \varphi)_{\mathbf{3}_S} (\tilde{\varphi} \varphi)_{\mathbf{3}_A} \big)_\mathbf{1} \,,
\label{eq:VPhi} 
\end{eqnarray}
where $f_i$ are real and $\tilde{\varphi}=(\varphi_1^*, \varphi_3^*, \varphi_2^*)$ also transforms as a $\mathbf{3}$ of $A_4$. Terms related to the antisymmetric combination $(\tilde{\varphi} \varphi)_{\mathbf{3}_A}$ are included due to the complex property of $\varphi$. After $\varphi$ gets the VEV $\langle \varphi \rangle = (1,0,0)^T v_\varphi/\sqrt{2}$ and $A_4$ is broken to $Z_3$, $\varphi_1$, $\varphi_2$ and $\varphi_3$ get masses with mass eigenvalues
\begin{eqnarray}
&&m_{\varphi_1}^2=2(f_1+f_3)v_\varphi^2\,, \quad 
m_{\varphi_2^{\prime}}^2,m_{\varphi_3^{\prime}}^2= \left(2 f_2 - 5 f_3 + f_4 \pm \sqrt{(2 f_2 + f_3 - f_4)^2 + 4 f_5^2}\right) \frac{v_\varphi^2}{4} \,.
\label{eq:flavon_mass_sub}
\end{eqnarray}
The mixing angle $\vartheta$ is given by $\cot2\vartheta=2f_5/(2f_2+f_3-f_4) $ \footnote{As mentioned in the last section, the convention $-45^\circ\leqslant\vartheta\leqslant 45^\circ$ is used. This is equivalent to that in \cite{PZ}, in which both $-90^\circ\leqslant\vartheta\leqslant 90^\circ$ and the mass ordering $m_{\varphi_2'}\leqslant m_{\varphi_3'}$ are required.}. The potential of $\chi$ can be obtained with the replacements $\varphi \to \chi$, and $f_i \to g_i$ in Eq.~\eqref{eq:VPhi}. In order to achieve the successful breaking of $A_4 \to Z_3$ and $Z_2$ in charged lepton and neutrino sectors, respectively, the following conditions must be satisfied: 
\begin{eqnarray}
&f_1+f_3>0\,,\quad
2f_2-5f_3+f_4>0\,,\quad
2(f_2-f_3)(f_4-3f_3)-f_5^2>0\,, \nonumber\\
&g_1+g_2>0\,,\quad
3g_3-6g_2+g_4>0\,,\quad
4(g_2-g_3)(3g_2-g_4)-g_5^2>0\,. 
\label{eq:stable_vacuum}
\end{eqnarray} 
%%%
The cross couplings between $\varphi$ and $\chi$ are given by
\begin{eqnarray}
\hspace{-7mm }
V(\varphi, \chi) &=& 2 \epsilon_1( \tilde{\varphi} \varphi )_\mathbf{1} ( \tilde{\chi} \chi )_\mathbf{1} + \big[ \epsilon_2 ( \tilde{\varphi} \varphi )_{\mathbf{1}''} ( \tilde{\chi} \chi )_{\mathbf{1}'} +\text{h.c.} \big]
+ 2 \epsilon_3 \big( ( \tilde{\varphi} \varphi )_{\mathbf{3}_S} ( \tilde{\chi} \chi )_{\mathbf{3}_S} \big)_{\mathbf{1}} \nonumber\\
&&+ 2 \epsilon_4 \big( ( \tilde{\varphi} \varphi )_{\mathbf{3}_A} ( \tilde{\chi} \chi )_{\mathbf{3}_A} \big)_{\mathbf{1}} + 2 \epsilon_5 \big( ( \tilde{\varphi} \varphi )_{\mathbf{3}_S} ( \tilde{\chi} \chi )_{\mathbf{3}_A} \big)_{\mathbf{1}} + 2 \epsilon_6 \big( ( \tilde{\varphi} \varphi )_{\mathbf{3}_A} ( \tilde{\chi} \chi )_{\mathbf{3}_S} \big)_{\mathbf{1}} \,,
\label{eq:Vmix2}
\end{eqnarray}
in which $\epsilon_2$ is complex and $\epsilon_1$, $\epsilon_3$, $\epsilon_4$, $\epsilon_5$ and $\epsilon_6$ are real parameters, which we assume to be small. Here, the $\epsilon_2$ term is the only one that modifies the VEV of $\varphi$ at first order. Taking into account its contribution of this term, we finally obtain the corrections to the VEVs, see Eq.~\eqref{eq:vev_shift}, 
\begin{eqnarray}
\epsilon_{\varphi2}=(1-\kappa)\epsilon_{\varphi} \,,\qquad
\epsilon_{\varphi3}=(1+\kappa)\epsilon_{\varphi}^* 
\end{eqnarray}
with 
\begin{eqnarray} 
\epsilon_{\varphi}=-\frac{ \epsilon_2 v_\chi^2 \big[(m_{\varphi_3'}^2 + m_{\varphi_2'}^2) - (m_{\varphi_3'}^2 - m_{\varphi_2'}^2) \sin2\vartheta \big]}{4 m_{\varphi_3'}^2 m_{\varphi_2'}^2} \,, \quad
\kappa=\frac{ (m_{\varphi_3'}^2 - m_{\varphi_2'}^2) \cos2\vartheta }{ (m_{\varphi_3'}^2 + m_{\varphi_2'}^2) - (m_{\varphi_3'}^2 - m_{\varphi_2'}^2) \sin2\vartheta} \,. 
\label{eq:vevs3}
\end{eqnarray} 
When the mixing between $\varphi_2$ and $\varphi_3$ is maximal, $\sin2\vartheta=\pm1$, $\kappa$ vanishes, and we get the same structure of the VEV shift as in Model I. Furthermore, $\epsilon_\varphi$ takes the value $- \epsilon_2 v_\chi^2/(2m_{\varphi_3'}^2)$  and $- \epsilon_2 v_\chi^2/(2m_{\varphi_2'}^2)$
for $\vartheta=45^\circ$ and $-45^\circ$, respectively. 

The Lagrangian terms led to lepton masses is also Eq.~\eqref{eq:Yukawa_coupling}, the same as Model I. Since the $\kappa$-related asymmetric correction is included in the VEV $\langle \varphi \rangle$, the expressions for the mixing parameters are modified, approximating to \cite{PZ}
\begin{eqnarray}
&&\sin\theta_{13}= |\epsilon_\varphi|\sqrt{2 \kappa^2\cos^2\theta_\varphi + 2 \sin^2 \theta_\varphi}\,,\nonumber\\
&&\sin\theta_{12}= \frac{1}{\sqrt{3}}\big[1-2|\epsilon_\varphi|\cos\theta_\varphi\big]\,,\nonumber\\
&&\sin\theta_{23}= \frac{1}{\sqrt{2}}\big[1+(1+\kappa)|\epsilon_\varphi| \cos\theta_\varphi \big] \,,\nonumber\\
&&\delta=\text{Arg} \left\{ \Big[ -i \sin\theta_\varphi - \kappa\cos\theta_\varphi \Big] \Big[ 1-i |\epsilon_\varphi| (2+\kappa) \sin\theta_\varphi \Big] \right\} \,.  
\label{eq:mixing_angles2}
\end{eqnarray}
The combination of $\kappa$ and the real part of $\epsilon_2$ provides new sources for $\theta_{13}$ and CP violation. It can induce sizable $\theta_{13}$ while not affecting CP conservation in some specific region of the parameter space \cite{PZ}. In the case of maximal mixing between $\varphi_2$ and $\varphi_3$, $\cos2\vartheta=0$, leading to $\kappa=0$, we recover $\theta_{13}=\sqrt{2}|\epsilon_\varphi \sin \theta_\varphi|$ and nearly maximal Dirac-type CP violation, the same result as in Model I.

\subsection{$Z_3$-breaking LFV charged lepton decays in Model I} 

We now discuss the $Z_3$-breaking charged LFV processes induced by the cross couplings between $\varphi$ and $\chi$ in Model I. 
The mixing and mass splitting of different components of $\varphi$ induced by $\epsilon_2$ can be expressed in terms of $\epsilon_{\varphi_1\varphi_2}=c\epsilon_\varphi$ and $\epsilon_{h_2 a_2}=\epsilon_\varphi$ with $c=(m_{\varphi_2}^2 + m_{\varphi_1}^2)/(m_{\varphi_2}^2 - m_{\varphi_1}^2)$. As all the $Z_3$-breaking effects are essentially dependent upon $\epsilon_\varphi$, we expect all the LFV charged lepton decay modes to be suppressed by $|\epsilon_\varphi|^2$, or equivalently, suppressed by $r^2+ s^2 = 2s_{13}^2+(1-\sqrt{3}s_{12})^2$. 

The effective 4-fermion operators for  $\tau^-\to \mu^+\mu^-e^-$ after integrating out $\varphi_1$ and $\varphi_2$ are given by 
\begin{eqnarray}
-2 \epsilon_\varphi\frac{m_\mu m_\tau}{v_\varphi^2} 
\left[ \frac{1 }{m_{\varphi_1}^2} (\overline{\mu}\mu)(\overline{e_L}\tau_R) +
\frac{ 1}{m_{\varphi_2}^2}  (\overline{\mu_R}\mu_L)(\overline{e_L}\tau_R) + \frac{ 1 }{m_{\varphi_2}^2}  (\overline{e_L}\mu_R)(\overline{\mu_L}\tau_R) \right] \,.
\label{eq:tau_decay1}
\end{eqnarray}
Here, we have considered all contributions of order $\epsilon_\varphi m_\mu m_\tau/v_\varphi^4$. They include those due to the mixing of left-handed charged leptons $e_L$, $\mu_L$, $\tau_L$ and to the mixing between flavons $\varphi_1$ and $\varphi_2$.
The effective 4-fermion interaction relevant for $\tau^-\to \mu^+\mu^-\mu^-$ and $\tau^-\to e^+ e^- \mu^- $ are given by
\begin{eqnarray} 
&& -2 \epsilon_\varphi^* \frac{m_\mu m_\tau}{v_\varphi^2} 
\left[ \frac{1}{m_{\varphi_1}^2}  (\overline{\mu}\mu)(\overline{\mu_L}\tau_R) +
\frac{1}{m_{\varphi_2}^2}  (\overline{\mu_R}\mu_L)(\overline{\mu_L}\tau_R) \right] \,,\nonumber\\
&& -2 \epsilon_\varphi^* \frac{m_\mu m_\tau}{v_\varphi^2} \frac{1}{m_{\varphi_2}^2} 
(\overline{\mu_R}e_L)(\overline{e_L}\tau_R)  \,, 
\label{eq:tau_decay2}
\end{eqnarray}
respectively. From Eqs.~\eqref{eq:tau_decay1} and \eqref{eq:tau_decay2}, we obtain simple relations of the non-zero effective coefficients of the 4-fermion interactions of these processes
\begin{eqnarray}
&&(C^{\mu\mu\mu\tau}_{LRLR})^* = C^{\mu\mu e\tau}_{LRLR} = -2 \epsilon_\varphi \frac{m_\mu m_\tau}{v_\varphi^2 m_{\varphi_1}^2} \,, \nonumber\\
&&(C^{\mu ee\tau}_{RLLR})^* = C^{e\mu\mu\tau}_{LRLR} = -2 \epsilon_\varphi \frac{m_\mu m_\tau}{v_\varphi^2 m_{\varphi_2}^2} \,, \nonumber\\
&&(C^{\mu\mu\mu\tau}_{RLLR})^* = C^{\mu\mu e\tau}_{RLLR} = (C^{\mu\mu\mu\tau}_{LRLR})^*+ (C^{\mu ee\tau}_{RLLR})^* \,,
\end{eqnarray}
and the other coefficients will not contribute to the decays at leading order. These simple relations result in sum rules for the branching ratios of the $\tau$ LFV decays
\begin{eqnarray}
& 2(B_{\mu^+\mu^-e^-}-2B_{\mu^+\mu^-\mu^-})^2+(5B_{e^+e^-\mu^-}+10B_{\mu^+\mu^-\mu^-}-6B_{\mu^+\mu^-e^-})B_{e^+e^-\mu^-}=0 \,,\nonumber\\
& B_{e^+e^-\mu^-} \approx 2 (r^2+s^2) \text{Br}(\tau^-\to \mu^+ e^- e^-) \,,
\label{eq:branching_sum_rule2}
\end{eqnarray}
where $B_{\mu^+\mu^-e^-}$, $B_{\mu^+\mu^-\mu^-}$, $B_{e^+e^-\mu^-}$ are branching ratios of $\tau^-\to \mu^+\mu^-e^-$, $\tau^-\to \mu^+\mu^-\mu^-$ and $\tau^-\to e^+ e^- \mu^- $, respectively. In the limit $m_{\varphi_1}\ll m_{\varphi_2}$, we get $B_{\mu^+\mu^-e^-}\approx 2 B_{\mu^+\mu^-\mu^-} \gg B_{e^+e^-\mu^-}$, and on the contrary, we have $B_{\mu^+\mu^-e^-} \approx 4 B_{\mu^+\mu^-\mu^-} \approx 2 B_{e^+e^-\mu^-}$. 

For the processes $\tau^-\to e^+ e^-e^-$ and $\mu^-\to e^+ e^-e^-$, coefficients for the related operators are
\begin{eqnarray}
C^{eee\tau}_{LRLR} = C^{eee\tau}_{RLLR} = -2 \epsilon_\varphi \frac{m_e m_\tau}{v_\varphi^2} \left( \frac{1}{m_{\varphi_1}^2} + \frac{1}{m_{\varphi_2}^2} \right) \,, \nonumber\\
C^{eee\mu}_{LRLR} = C^{eee\mu}_{RLLR} = -2 \epsilon_\varphi^* \frac{m_e m_\mu}{v_\varphi^2} \left( \frac{1}{m_{\varphi_1}^2} + \frac{1}{m_{\varphi_2}^2} \right) \,,
\end{eqnarray}
both suppressed by the electron mass. Taking $v_\varphi\sim m_{\varphi_1} \sim m_{\varphi_2}$ around the electroweak scale, we obtain branching ratios smaller than $10^{-16}$ and $10^{-18}$, respectively, far below the current experimental upper limit. The $Z_3$-breaking terms will also contribute to the channels $\tau^-\to \mu^+e^- e^- $ and $\tau^- \to e^+ \mu^- \mu^- $. They are subleading, generically suppressed by $\epsilon_\varphi$, and will not be considered here. 

%%%
The radiative decay $l_1^-\to l_2^- \gamma$ is allowed after the flavon cross couplings are included. We calculate them in detail in the Appendix and we show the results here. 
For $\tau^- \to e^- \gamma$, there are two main contributions: one is due to the mixing between $e_L$ and $\tau_L$ and the other to the mixing between $Z_3$-invariant and $Z_3$-covariant flavons $\varphi_1$ and $\varphi_2$. After the transformation in Eqs.~\eqref{eq:lepton_shift} and \eqref{eq:flavon_shift}, we obtain the corresponding operators \footnote{We just keep the terms directly to the process. Some  Hermitian terms not relevant to $l_1^-\to l_2^- \gamma$ are not given here.}: 
\begin{eqnarray}
\frac{m_\tau}{v_\varphi}\left[
 \overline{e_L} \tau_R \left( \varphi_2 - \tilde{\epsilon}_\varphi \varphi_1 \right) + \overline{\tau} \tau \left(\varphi_1 +  \tilde{\epsilon}_\varphi \varphi_2^*\right) 
\right] \,,
\label{eq:Ltau2egamma}
\end{eqnarray}
where $\tilde{\epsilon}_\varphi = (1+c)\epsilon_\varphi = {2 \epsilon_\varphi m_{\varphi_2}^2 }/{(m_{\varphi_2}^2- m_{\varphi_1}^2)} $. 
For $\tau^-\to \mu^- \gamma$, the main sources are the mixing between $\mu_L$ and $\tau_L$ and the one between $\varphi_1$ and $\varphi_2$. Operators related to these contributions are 
\begin{eqnarray}
\frac{m_\tau}{v_\varphi}\left[
 \overline{\mu_L} \tau_R \left( \varphi_2^* - \tilde{\epsilon}_\varphi^* \varphi_1 \right) + \overline{\tau} \tau \left(\varphi_1 +  \tilde{\epsilon}_\varphi^* \varphi_2\right) 
\right]
\label{eq:Ltau2mugamma}
\end{eqnarray}
For $\mu^-\to e^- \gamma$, the main contributions come from the mixing between $\mu_L$ and $\tau_L$, the mixing between $e_L$ and $\tau_L$, and the mass splitting between $h_2$ and $a_2$. $\varphi_1$ is not involved in this channel. After going into the charged lepton mass basis and replacing $\varphi_2$ by $e^{-i\theta_\varphi/2}(h_2+ia_2)/\sqrt{2}$ in Eq.~\eqref{eq:varphi_real}, the relevant operators are  
\begin{eqnarray}
&&\hspace{10mm}\frac{e^{-i\frac{\theta_\varphi}{2}}}{\sqrt{2}} \frac{m_\tau}{v_\varphi} \overline{e_L} \tau_R 
\Big[\left( 1- |\epsilon_\varphi|\right) h_2 + i\left(1 + |\epsilon_\varphi| \right)  a_2 \Big] \nonumber\\
&+&\frac{e^{-i\frac{\theta_\varphi}{2}}}{\sqrt{2}}\Big[\frac{m_\tau}{v_\varphi} \overline{\tau_R} \mu_L 
+\frac{m_\mu}{v_\varphi} \overline{\tau_L} \mu_R \Big]
\Big[\left( 1 + |\epsilon_\varphi|\right) h_2 + i\left(1 - |\epsilon_\varphi| \right)  a_2 \Big] \,.
\label{eq:Lmu2egamma}
\end{eqnarray}
In the above equation, we keep the term $\overline{\tau_R} \mu_L$ at the ${m_\tau}/{v_\varphi}$ level but keep $\overline{\tau_L} \mu_R$ at the ${m_\mu}/{v_\varphi}$ one. The reason is that the former contribution is proportional to $m_\mu$ and the latter contribution is proportional to $m_\tau$, as shown in Eq.~\eqref{eq:Integral} of the Appendix. 

Starting from the effective couplings in Eqs.~\eqref{eq:Ltau2egamma}, \eqref{eq:Ltau2mugamma} and \eqref{eq:Lmu2egamma}, and following the calculation in the Appendix, we finally arrive at 
\begin{eqnarray}
&&
A_{R}^{e\tau}= \frac{ie \tilde{\epsilon}_\varphi m_\tau}{(4\pi)^2 v_\varphi^2} \left[ F(\varphi_2) - F(\varphi_1) \right] \,,\nonumber\\
&&
A_{R}^{\mu\tau}= \frac{ie \tilde{\epsilon}_\varphi^* m_\tau}{(4\pi)^2 v_\varphi^2} \left[ F(\varphi_2) - F(\varphi_1) \right] \,,\nonumber\\
&&
A_{R}^{e\mu}= \frac{-2ie \epsilon_\varphi^* m_\mu}{(4\pi)^2 v_\varphi^2} \frac{m_\tau^2}{m_{\varphi_2}^2} \Big( \log\frac{m^2_\tau}{m_{\varphi_2}^2}+\frac{7}{3} \Big) \,,
\label{eq:AR}
\end{eqnarray}
where 
\begin{eqnarray}
F(\varphi_i)= \frac{m_\tau^2}{m_{\varphi_i}^2} \Big(\log\frac{m^2_\tau}{m_{\varphi_i}^2}+\frac{4}{3} \Big)\,.
\end{eqnarray}
We emphasise that we have taken account of all three contributions from the mixing of charged leptons, the mixing between $Z_3$-invariant and $Z_3$-covariant flavons $\varphi_1$ and $\varphi_2$, and the mass splitting between two components of the $Z_3$-covariant flavon $\varphi_2$, as shown in Eqs.~\eqref{eq:lepton_shift}, \eqref{eq:flavon_shift}, and \eqref{eq:real_masses}, respectively. 
As mentioned in the last subsection, we prove in the Appendix that $A_L^{l_2l_1} \ll A_R^{l_2l_1}$ is satisfied in all three channels, as imposed by the structure of charged lepton mass matrix. Here, one more sum rule for branching ratios is satisfied: 
\begin{eqnarray}
\text{Br}(\tau^- \to e^- \gamma) \approx \text{Br}(\tau^-\to \mu^- \gamma) \,.
\label{eq:branching_sum_rule3}
\end{eqnarray}

\begin{figure}[h!]
\begin{center}
\includegraphics[width=0.6\textwidth]{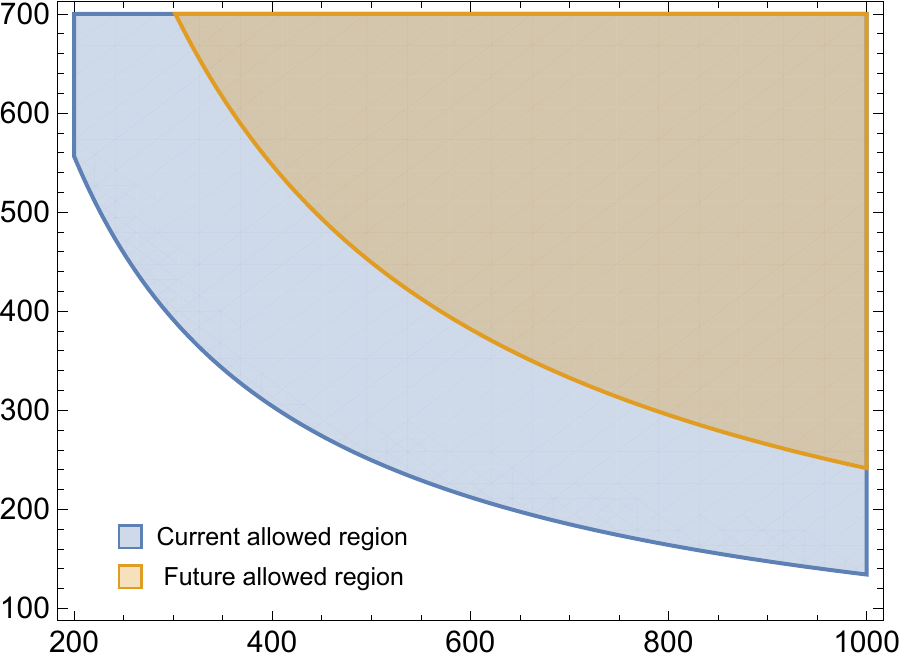}
\vspace{-.5cm} 
\caption{The current and near future constraints on Model I from the $\mu^-\to e^- \gamma$ experiments. $|\epsilon_\varphi|$ is fixed at 0.1 for generating the reactor angle $\theta_{13}$. The current constraint of the MEG experiment is set to be $\text{Br}(\mu^-\to e^-\gamma)<4.2\times 10^{-13}$ \cite{MEG}, and the future constraint of MEG II is set to be $\text{Br}(\mu^- \to e^- \gamma)<4\times 10^{-14}$ \cite{MEG2}. }
\label{fig:muI}
\end{center}
\end{figure}

Taken the scale of the flavour symmetry to be at the electroweak scale, the branching ratio of $\tau^- \to e^- \gamma$ and $\tau^- \to \mu^- \gamma$ are in general less than $10^{-11}$, at least 3 orders of magnitude below the current experimental upper limits. 
The $\mu^-\to e^- \gamma$ channel has been measured most precisely and gives the strongest constraint on the model. In Fig. \ref{fig:muI}, we show regions of $v_\varphi$ and $m_{\varphi_2}$ allowed by current experiments and testable at the near future experiments. The current upper limit of the branching ratio is $4.2\times 10^{-13}$, measured by the MEG experiment \cite{MEG}. By fixing the flavon VEV $v_\varphi=v/\sqrt{2},\, v,\, 2v=175,\, 246,\, 492$ GeV, we obtain the lower limit of the $\varphi_2$ mass: $m_{\varphi_2}>700,\, 500,\, 200$ GeV, respectively. The MEG II experiment will reach the upper limit of the branching ratio to $4\times 10^{-14}$ in the near future \cite{MEG2}. This experiment has the potential to prove the scale of flavour symmetry around electroweak scale or push it to a higher scale. 

%%%%%%%%%%%%%%%%%
%%%%%%%%%%%%%%%%%

\subsection{$Z_3$-breaking LFV charged lepton decays in Model II \label{sec:pII}} 

Flavon cross couplings shift the $Z_3$-preserving VEVs and open the $Z_3$-breaking LFV channels. Similar to Model I, the shift of flavon VEVs in Model II results in three contributions to the $Z_3$-breaking processes. They are
\begin{itemize} 
\item the mixing between left-handed charged leptons. To get the Lagrangian in the charged lepton mass eigenstates, we do the transformation 
\begin{eqnarray}
e_L &\to& e_L+(1+\kappa)\epsilon_\varphi^* \mu_L + (1-\kappa)\epsilon_\varphi \tau_L \,,\nonumber\\
\mu_L &\to& \mu_L+(1+\kappa)\epsilon_\varphi^* \tau_L - (1-\kappa)\epsilon_\varphi e_L \,,\nonumber\\
\tau_L &\to& \tau_L-(1+\kappa)\epsilon_\varphi^* e_L - (1-\kappa)\epsilon_\varphi \mu_L \,.
\label{eq:lepton_shift2} 
\end{eqnarray}

\item the mixing between $Z_3$-invariant flavon $h_1$ and $Z_3$-covariant flavons $\varphi_2'$, $\varphi_3'$. It leads to the following transformation to the flavon mass eigenstates:  
\begin{eqnarray}
&&h_1\to h_1+ c_{12} \left(\epsilon_\varphi^* \varphi_2' + \epsilon_\varphi \varphi_2^{\prime*}\right) + c_{13} \left(\epsilon_\varphi \varphi_3' + \epsilon_\varphi^* \varphi_3^{\prime*}\right)  \,, \nonumber\\ 
&&\varphi_2'\to \varphi_2' - c_{12} \epsilon_\varphi h_1 + c_{23} |\epsilon_\varphi|  \varphi_3^{\prime*}  \,,\nonumber\\
&&\varphi_3'\to \varphi_3' - c_{13} \epsilon_\varphi^* h_1 - c_{23} |\epsilon_\varphi|  \varphi_2^{\prime*}  \,,
\label{eq:flavon_shift2}
\end{eqnarray}
in which 
\begin{eqnarray}
&&c_{12}=\frac{m_{\varphi_2'}^2+m_{\varphi_1}^2}{m_{\varphi_2'}^2-m_{\varphi_1}^2} \frac{m_{\varphi_3'}^2 \sqrt{2-2\sin2\vartheta}}{(m_{\varphi_3'}^2 + m_{\varphi_2'}^2) - (m_{\varphi_3'}^2 - m_{\varphi_2'}^2) \sin2\vartheta} \,,\nonumber\\
&&c_{13}=\frac{m_{\varphi_3'}^2+m_{\varphi_1}^2}{m_{\varphi_3'}^2-m_{\varphi_1}^2} \frac{m_{\varphi_2'}^2 \sqrt{2+2\sin2\vartheta}}{(m_{\varphi_3'}^2 + m_{\varphi_2'}^2) - (m_{\varphi_3'}^2 - m_{\varphi_2'}^2) \sin2\vartheta} \,,\nonumber\\
&&c_{23}=\frac{2\big[(m_{\varphi_3'}^2+m_{\varphi_2'}^2)^2 - 3(m_{\varphi_3'}^2-m_{\varphi_2'}^2)^2 \cos4\vartheta\big]\cos2\vartheta}{(m_{\varphi_3'}^2-m_{\varphi_2'}^2)\big[(m_{\varphi_3'}^2 + m_{\varphi_2'}^2) - (m_{\varphi_3'}^2 - m_{\varphi_2'}^2) \sin2\vartheta\big]}  \,.
\end{eqnarray}

\item the mass splitting between $h_2'$ and $a_2'$, and that between $h_3'$ and $a_3'$, in which $h_2'$, $a_2'$, $h_3'$ and $a_3'$ are mass eigenstates of $\varphi_2'$ and $\varphi_3'$ given by
\begin{eqnarray}
h_2=\frac{1}{\sqrt{2}} \Big(\varphi_2 \exp(i\frac{\theta_{\varphi}}{2}) + \varphi_2^* \exp(-i\frac{\theta_{\varphi}}{2})\Big)\,,\quad &&
h_3=\frac{1}{\sqrt{2}} \Big(\varphi_3 \exp(i\frac{\theta_{\varphi}}{2}) + \varphi_3^* \exp(-i\frac{\theta_{\varphi}}{2})\Big) \,,\nonumber\\
a_2=\frac{-i}{\sqrt{2}} \Big(\varphi_2 \exp(i\frac{\theta_{\varphi}}{2}) - \varphi_2^* \exp(-i\frac{\theta_{\varphi}}{2})\Big)\,,\quad &&
a_3=\frac{i}{\sqrt{2}} \Big(\varphi_3 \exp(i\frac{\theta_{\varphi}}{2}) - \varphi_3^* \exp(-i\frac{\theta_{\varphi}}{3})\Big)
\label{eq:varphi_real}
\end{eqnarray} 
The corrected masses of them are respectively given by
\begin{eqnarray} 
m_{h_2',a_2'}^2=m_{\varphi_2'}^2(1\pm c_{22}|\epsilon_\varphi|)\,, \quad
m_{h_3',a_3'}^2=m_{\varphi_3'}^2(1\pm c_{33}|\epsilon_\varphi|)\,,
\end{eqnarray} 
with 
\begin{eqnarray} 
&&c_{22} = \frac{(1-\sin2\vartheta)\big[ (m_{\varphi_3'}^2 + m_{\varphi_2'}^2)^2 - (m_{\varphi_3'}^2 - m_{\varphi_2'}^2)^2(1+6 (1+\sin2\vartheta) \sin2\vartheta) \big]}{2m_{\varphi_2'}^2 \big[(m_{\varphi_3'}^2 + m_{\varphi_2'}^2) - (m_{\varphi_3'}^2 - m_{\varphi_2'}^2) \sin2\vartheta\big]} \,,\nonumber\\
&&c_{33} = \frac{(1+\sin2\vartheta)\big[ (m_{\varphi_3'}^2 + m_{\varphi_2'}^2)^2 - (m_{\varphi_3'}^2 - m_{\varphi_2'}^2)^2(1-6 (1-\sin2\vartheta) \sin2\vartheta) \big]}{2m_{\varphi_3'}^2 \big[(m_{\varphi_3'}^2 + m_{\varphi_2'}^2) - (m_{\varphi_3'}^2 - m_{\varphi_2'}^2) \sin2\vartheta\big]} \,.
\end{eqnarray} 
The terms $1\pm c_{22} |\epsilon_\varphi|$ and $1\pm c_{33} |\epsilon_\varphi|$ must be positive to stabilise the vacuum. 
\end{itemize}
Here, we would like to mention two special cases, $\vartheta=\pm 45^\circ$: 
\begin{itemize}

\item In the case $\vartheta= 45^\circ$, we have $c_{12}$, $c_{23}$ and $c_{22}$ vanish, $c_{13}$ takes the same value as $c$ in Model I, and $c_{33}=2$. In this case,  $h_1$ and $\varphi_3^{\prime*}$ (as well as $h_3'$ and $a_3'$) are identical with $\varphi_1$, $\varphi_2$ (as well as $h_2$ and $a_2$) in Model I, respectively. There is no $\epsilon_\varphi$-induced mixing between $\varphi_2'$ and $h_1$, nor that between $\varphi_2'$ and $\varphi_3^{\prime*}$.  

\item In the case $\vartheta= - 45^\circ$, we have $c_{13}$, $c_{23}$ and $c_{33}$ vanish, $c_{12}$ takes the same value as $c$ in Model I, and $c_{22}=2$. Therefore, $h_1$ and $\varphi_2'$ (as well as $h_2'$ and $a_2'$) are identical with $\varphi_1$, $\varphi_2$ (as well as $h_2$ and $a_2$) in Model I, respectively, and there is no $\epsilon_\varphi$-induced mixing between $\varphi_3^{\prime*}$ and $h_1$, nor that between $\varphi_3^{\prime*}$ and $\varphi_2'$.

\end{itemize}

After considering the three effects of $Z_3$-breaking, we can repeat the procedure as in Model I to calculate the LFV charged lepton decays and derive the 4-fermion interactions for $\tau^-\to \mu^+ \mu^- e^-$, $\tau^-\to \mu^+\mu^-\mu^-$, $\tau^-\to e^+e^-\mu^-$, $\tau^-\to e^+e^-e^-$ and $\mu^-\to e^+e^-e^- $. Here, we list the coefficients that will contribute to the decays at leading order: 
\begin{eqnarray}
C^{\mu\mu e\tau}_{LRLR} &=& - \epsilon_\varphi\frac{m_\mu m_\tau}{v_\varphi^2} \left[
\frac{1+\kappa}{m_{\varphi_1}^2} + \sqrt{2} c_{12} c_\vartheta \frac{\Delta m^2_{\varphi_2' \varphi_1}}{m_{\varphi_1}^2 m_{\varphi_2'}^2} + \sqrt{2} c_{13} s_\vartheta \frac{\Delta m^2_{\varphi_3' \varphi_1}}{m_{\varphi_1}^2 m_{\varphi_3'}^2} + (1+\kappa)\sin2\vartheta \Big(\frac{1}{m_{\varphi_2'}^2} - \frac{1}{m_{\varphi_3'}^2}\Big) \right]\,,\nonumber\\
C^{\mu\mu e\tau}_{RLLR} &=& - \epsilon_\varphi\frac{m_\mu m_\tau}{v_\varphi^2} \left[
\frac{1+\kappa}{m_{\varphi_1}^2} + \sqrt{2} c_{12} c_\vartheta \frac{\Delta m^2_{\varphi_2' \varphi_1}}{m_{\varphi_1}^2 m_{\varphi_2'}^2} + \sqrt{2} c_{13} s_\vartheta \frac{\Delta m^2_{\varphi_3' \varphi_1}}{m_{\varphi_1}^2 m_{\varphi_3'}^2}
+2 (1-\kappa) \Big( \frac{c^2_\vartheta}{m_{\varphi_2'}^2} + \frac{s^2_\vartheta}{m_{\varphi_3'}^2}\Big) \right]\,,\nonumber\\
C^{e\mu\mu\tau}_{LRLR} &=& - \epsilon_\varphi\frac{m_\mu m_\tau}{v_\varphi^2} \left[ 2 \Big( \frac{c_{22}s^2_\vartheta}{m_{\varphi_2'}^2} + \frac{c_{33}c^2_\vartheta}{m_{\varphi_3'}^2} \Big)  \right] \,; \nonumber\\
%%%
C^{\mu\mu\mu\tau}_{LRLR} &=& - \epsilon_\varphi^* \frac{m_\mu m_\tau}{v_\varphi^2} \left[
\frac{1+\kappa}{m_{\varphi_1}^2} - \sqrt{2} c_{12} s_\vartheta \frac{\Delta m^2_{\varphi_2' \varphi_1}}{m_{\varphi_1}^2 m_{\varphi_2'}^2} + \sqrt{2} c_{13} c_\vartheta \frac{\Delta m^2_{\varphi_3' \varphi_1}}{m_{\varphi_1}^2 m_{\varphi_3'}^2} + (1-\kappa)\sin2\vartheta \Big( \frac{1}{m_{\varphi_3'}^2} - \frac{1}{m_{\varphi_2'}^2} \Big) \right]\,, \nonumber\\
C^{\mu\mu\mu\tau}_{LRLR} &=& - \epsilon_\varphi^* \frac{m_\mu m_\tau}{v_\varphi^2} \left[
\frac{1+\kappa}{m_{\varphi_1}^2} - \sqrt{2} c_{12} s_\vartheta \frac{\Delta m^2_{\varphi_2' \varphi_1}}{m_{\varphi_1}^2 m_{\varphi_2'}^2} + \sqrt{2} c_{13} c_\vartheta \frac{\Delta m^2_{\varphi_3' \varphi_1}}{m_{\varphi_1}^2 m_{\varphi_3'}^2}
-2 (1+\kappa) \Big( \frac{s^2_\vartheta}{m_{\varphi_2'}^2} + \frac{c^2_\vartheta}{m_{\varphi_3'}^2} \Big)  \right] \,;\nonumber\\
%%%
C^{\mu e e\tau}_{RLLR} &=&-\epsilon_\varphi^* \frac{m_\mu m_\tau}{v_\varphi^2} 
\left[ \frac{2+2\kappa \cos\vartheta-c_{22}\sin2\vartheta}{m_{\varphi_2'}^2}+
\frac{2-2\kappa \cos\vartheta+c_{33} \sin2\vartheta}{m_{\varphi_3'}^2}  \right] (\overline{\mu_R}e_L) (\overline{e_L}\tau_R)  \,;\nonumber\\
%%%
C^{e e e\tau}_{LRLR} &=&- \epsilon_\varphi \frac{m_e m_\tau}{v_\varphi^2} \left[
\frac{1+\kappa}{m_{\varphi_1}^2} - \sqrt{2} c_{12} c_\vartheta \frac{\Delta m^2_{\varphi_2' \varphi_1}}{m_{\varphi_1}^2 m_{\varphi_2'}^2} - \sqrt{2} c_{13} s_\vartheta \frac{\Delta m^2_{\varphi_3' \varphi_1}}{m_{\varphi_1}^2 m_{\varphi_3'}^2} - (1+\kappa) \sin2\vartheta \Big( \frac{1}{m_{\varphi_2'}^2} - \frac{1}{m_{\varphi_3'}^2} \Big) \right]\,, \nonumber\\
C^{e e e\tau}_{RLLR} &=&- \epsilon_\varphi \frac{m_e m_\tau}{v_\varphi^2} \left[
\frac{1+\kappa}{m_{\varphi_1}^2} - \sqrt{2} c_{12} c_\vartheta \frac{\Delta m^2_{\varphi_2' \varphi_1}}{m_{\varphi_1}^2 m_{\varphi_2'}^2} - \sqrt{2} c_{13} s_\vartheta \frac{\Delta m^2_{\varphi_3' \varphi_1}}{m_{\varphi_1}^2 m_{\varphi_3'}^2} 
+2 (1-\kappa) \Big( \frac{s^2_\vartheta}{m_{\varphi_2'}^2} + \frac{c^2_\vartheta}{m_{\varphi_3'}^2} \Big) \right] \,; \nonumber\\
%%%
C^{e e e\mu}_{LRLR} &=& - \epsilon_\varphi^* \frac{m_e m_\mu}{v_\varphi^2} \left[
\frac{1-\kappa}{m_{\varphi_1}^2} - \sqrt{2} c_{12} s_\vartheta \frac{\Delta m^2_{\varphi_2' \varphi_1}}{m_{\varphi_1}^2 m_{\varphi_2'}^2} + \sqrt{2} c_{13} c_\vartheta \frac{\Delta m^2_{\varphi_3' \varphi_1}}{m_{\varphi_1}^2 m_{\varphi_3'}^2} - (1-\kappa) \sin2\vartheta \Big( \frac{1}{m_{\varphi_2'}^2} - \frac{1}{m_{\varphi_3'}^2} \Big) \right] \,, \nonumber\\
C^{e e e\mu}_{RLLR} &=& - \epsilon_\varphi^* \frac{m_e m_\mu}{v_\varphi^2} \left[
\frac{1-\kappa}{m_{\varphi_1}^2} - \sqrt{2} c_{12} s_\vartheta \frac{\Delta m^2_{\varphi_2' \varphi_1}}{m_{\varphi_1}^2 m_{\varphi_2'}^2} + \sqrt{2} c_{13} c_\vartheta \frac{\Delta m^2_{\varphi_3' \varphi_1}}{m_{\varphi_1}^2 m_{\varphi_3'}^2}+2 (1+\kappa) \Big( \frac{s^2_\vartheta}{m_{\varphi_2'}^2} + \frac{c^2_\vartheta}{m_{\varphi_3'}^2} \Big) \right] \,, 
\end{eqnarray} 
respectively. Here, $\Delta m^2_{\varphi_2' \varphi_1} \equiv m^2_{\varphi_2'} - m_{\varphi_1}^2$ and $\Delta m^2_{\varphi_3' \varphi_1} \equiv m^2_{\varphi_3'} - m_{\varphi_1}^2$.
%%%%%%%%%%%%%%%%%%%%%%%%%%%%%%%%%%%%
All these channels are suppressed by both $\epsilon_\varphi$ and charged lepton masses strongly dependent upon the mixing between the two $Z_3$-covariant flavons $\varphi_2$ and $\varphi_3$. In addition, the last two channels $\tau^-\to e^+e^-e^-$ and $\mu^-\to e^+e^- e^-$ are highly suppressed by the electron mass. From the above effective interactions, one can directly obtain branching ratios for these channels, which are of the same orders of magnitude as those in Model I.  

%Here we consider the simplification of the above formulae for different values of $\vartheta$. 

Finally, we discuss the modification to the branching ratio of $l_1^-\to l_2^- \gamma$. The coefficients $A_R^{l_2l_1}$ corresponding to $\tau^-\to e^-\gamma$, $\tau^-\to \mu^-\gamma$ and $\mu^-\to e^-\gamma$ are respectively given by 
\begin{eqnarray}
A_R^{e\tau}&=&\frac{ie \epsilon_\varphi m_\tau}{(4\pi)^2 v_\varphi^2} \left[ -(1+\kappa+\sqrt{2}c_{12}c_\vartheta+\sqrt{2}c_{13}s_\vartheta) F(h_1) \right.\nonumber\\
&&+
\sqrt{2}c_\vartheta (c_{12}-\sqrt{2}s_\vartheta (1+\kappa) ) F(\varphi_2')  
+ \sqrt{2}s_\vartheta (c_{13}+\sqrt{2}c_\vartheta (1+\kappa) ) F(\varphi_3') \nonumber\\
&&  \left.
- \frac{(2c_\vartheta s_\vartheta  (1+\kappa)  + 2 c_\vartheta^2 (1-\kappa) ) m_\tau^2}{6(4\pi)^2 m_{\varphi_2'}^2} 
+ \frac{(2c_\vartheta s_\vartheta  (1+\kappa) - 2 s_\vartheta^2 (1-\kappa) ) m_\tau^2}{6(4\pi)^2 m_{\varphi_3'}^2}\right]\,,\nonumber\\
%%%%%%%%%
A_R^{\mu\tau}&=&\frac{ie \epsilon_\varphi^* m_\tau}{(4\pi)^2 v_\varphi^2} \left[ -(1-\kappa-\sqrt{2}c_{12}s_\vartheta+\sqrt{2}c_{13}c_\vartheta) F(h_1) \right.\nonumber\\ 
&&-
\sqrt{2}s_\vartheta (c_{12}+\sqrt{2}c_\vartheta (1-\kappa) ) F(\varphi_2') + \sqrt{2}c_\vartheta (c_{13}+\sqrt{2}s_\vartheta (1-\kappa) ) F(\varphi_3') \nonumber\\ 
&& \left. - \frac{(2c_\vartheta s_\vartheta (1-\kappa)  + 2 s_\vartheta^2 (1+\kappa) ) m_\tau^2}{6(4\pi)^2 m_{\varphi_2'}^2} 
+ \frac{(2c_\vartheta s_\vartheta (1-\kappa)  - 2 c_\vartheta^2 (1+\kappa) ) m_\tau^2}{6(4\pi)^2 m_{\varphi_3'}^2}\right]\,,\nonumber\\
%%%%%%%%%
A_{R}^{e\mu}&=& \frac{-2ie \epsilon_\varphi^* m_\mu}{(4\pi)^2 v_\varphi^2} \left\{ \frac{m_\tau^2}{m_{\varphi_2'}^2} \left[ c_{22} c_\vartheta^2 \Big( \log\frac{m^2_\tau}{m_{\varphi_2'}^2}+\frac{5}{2} \Big) + \frac{1}{12}(c_{22} \sin2\vartheta + 2 \cos2\vartheta + 2\kappa) \right] \right.  \nonumber\\
&&\hspace{18.3mm}
\left.+\frac{m_\tau^2}{m_{\varphi_3'}^2} \left[ c_{33} c_\vartheta^2 \Big( \log\frac{m^2_\tau}{m_{\varphi_3'}^2}+\frac{5}{2} \Big) - \frac{1}{12}(c_{33} \sin2\vartheta + 2 \cos2\vartheta + 2\kappa) \right] \right\} . 
\label{eq:AR2}
\end{eqnarray}
Here, the analytical result of $\mu^-\to e^- \gamma$ is only valid in the case of small mass difference  between ${h_2'}$ and ${a_2'}$ and that between ${h_3'}$ and ${a_3'}$, i.e., $|c_{22}|, |c_{33}|\lesssim 2$. 
Again, $\varphi_1$ does not affect $\mu^-\to e^- \gamma$. 
We recover the results of Model I in the limit $\vartheta\to 45^\circ$ and $m_{\varphi_2'}^2\to \infty$, or in the limit $\vartheta\to -45^\circ$ and $m_{\varphi_3'}^2\to \infty$. 

%%%%%%%%%%%%%%%%%
\begin{figure}[h!]
\begin{center}
\includegraphics[width=0.9\textwidth]{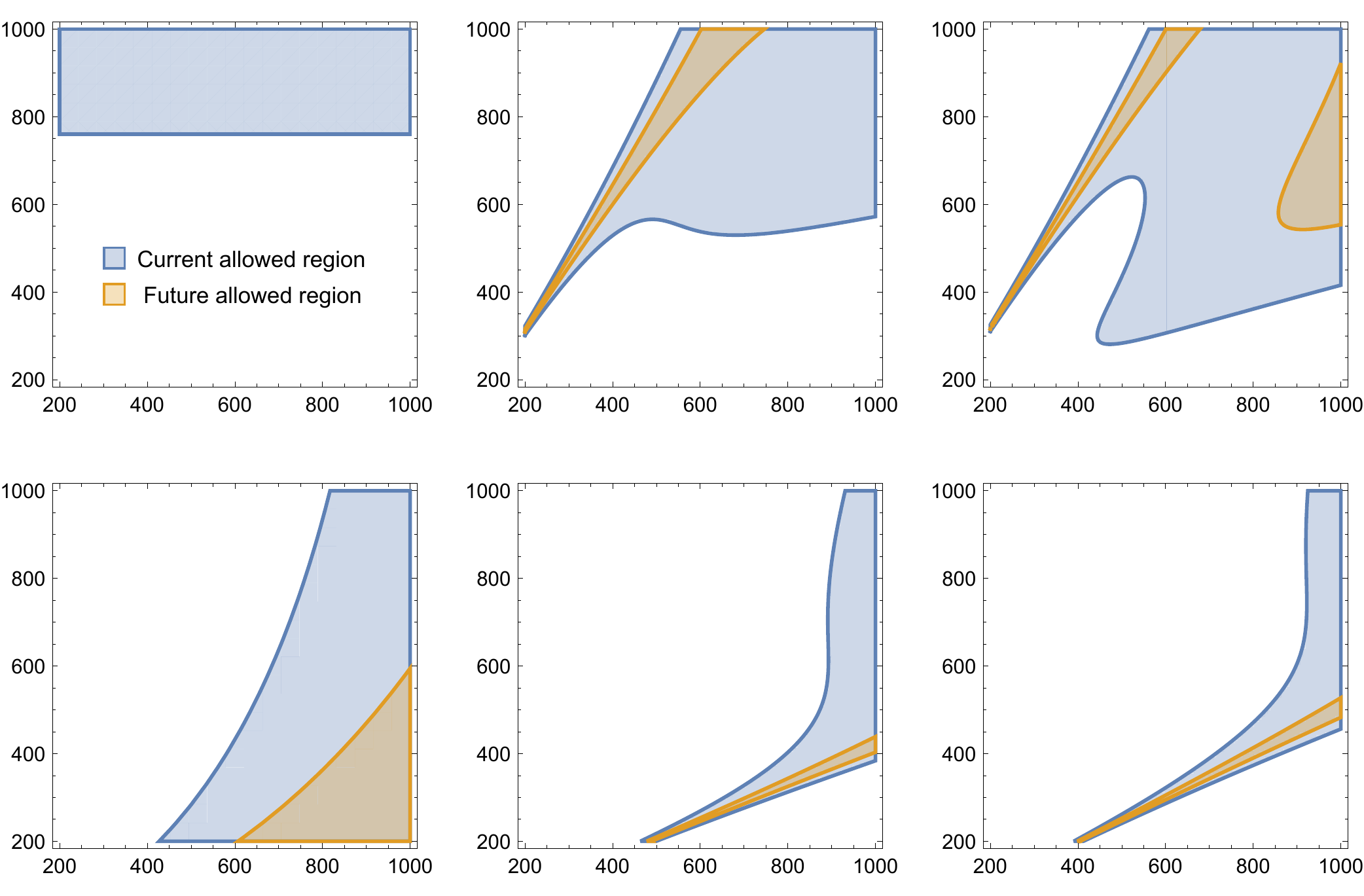}
\vspace{-.5cm} 
\caption{The current and near future constraints on the mass parameters $m_{\varphi_2'}$ and $m_{\varphi_3'}$ from the $\mu\to e \gamma$ experiments. $v_\varphi=175$ GeV is assumed. }
\label{fig:muII}
\end{center}
\end{figure}
%%%%%%%%%%%%%%%%%

We perform the numerical analysis to show how Model II is constrained by current experiments. $\tau^-\to e^-\gamma$ and $\tau^-\to \mu^-\gamma$ are safe for current experimental limit by assuming the flavour symmetry scale around the electroweak one. Also for Model II,
the strongest constraint is also from the $\mu^-\to e^-\gamma$ searches. In order to avoid the situation where the perturbation theory is not valid, we directly apply Eqs.~\eqref{eq:Feynman_radiative2} and \eqref{eq:AR22} in the Appendix into our numerical calculation. 
We fix $v_\varphi=v/\sqrt{2}=175$ GeV, vary the mixing angle $\vartheta$ and show the allowed regions of the mass parameters $m_{\varphi_2'}$ and $m_{\varphi_3'}$ by current and the expected future experiments in Fig. \ref{fig:muII}. Some comments follow: 
\begin{itemize} 

\item Masses $m_{\varphi_2'}$ and $m_{\varphi_3'}$ as low as $200$ to $400$ GeV are still allowed for some values of $\vartheta$, much lower than the mass $m_{\varphi_2}$ in Model I. This corresponds to the cancellation of contributions of $\varphi_2$ and $\varphi_3$. 

\item The mass eigenvalues of $h_2'$, $a_2'$ and $h_3'$, $a_3'$ in general deviate from $m_{\varphi_2'}$ and $m_{\varphi_3'}$, and the relative deviations are characterisd by $\pm c_{22}|\epsilon_\varphi|$ and $\pm c_{33}|\epsilon_\varphi|$, respectively. Numerically, we have checked that $|c_{22}|, |c_{33}| \lesssim 2$ hold in most of the allowed parameter space in Fig. \ref{fig:muII}. Thus, the deviations of $m_{h_2'}$, $m_{a_2'}$ from $m_{\varphi_2'}$, as well as those of $m_{h_3'}$, $m_{a_3'}$ from $m_{\varphi_3'}$, are in general small, and we can treat $m_{\varphi_2'}$ and $m_{\varphi_3'}$ as their masses at leading order. 

\item We have also checked that in all the allowed parameter space, $|c_{22}\epsilon_\varphi|, \, |c_{33}\epsilon_\varphi| < 1$. In other words, positive masses $m_{h_2'}^2,\, m_{a_2'}^2,\, m_{h_3'}^2,\, m_{a_3'}^2>0$ are guaranteed. 

\item For $\vartheta=45^\circ$, $c_{22}$ vanishes, no mass deviation between $h_2'$ and $a_2'$ arises, and the contributions of $h_2'$ and $a_2'$ cancel with each other. Thus, there is no constraint on $m_{\varphi_2'}$. The current experimental upper limit allows the mass of $\varphi_3'$ to be larger than 700 GeV. In the future, a mass smaller than 1 TeV would be ruled out. This is consistent with the results of $\varphi_2$ in Model I in Fig. \ref{fig:muI}. We have also checked that for $\vartheta=-45^\circ$, there will be no constraint on $m_{\varphi_3'}$ and the upper limit of the $\varphi_2'$ mass is also around 700 GeV, which is not shown in Fig. \ref{fig:muII}. 

\end{itemize}

\section{Conclusion}

Varies of flavour models with discrete flavour symmetries have been proposed to understand the mystery of lepton flavour mixing. Essential ingredients are flavon fields which couple among themselves and with leptons. The flavon potential generates special vacuum expectation values for flavons and trigger flavour symmetry breaking. And the couplings with leptons are responsible for Yukawa couplings with special flavour structures after the flavons get VEVs. These couplings will unavoidably contribute to other lepton-flavour-violating processes beyond neutrino oscillations. In this paper, we discuss the LFV decays of charged leptons induced by these couplings. 

All charged LFV processes have the same origin as leptonic flavour mixing, as they originate from the effective couplings of flavons and couplings between flavons and leptons. For definiteness, we assume that the flavour symmetry is $A_4$ and lepton flavour mixing is tri-bimaximal at leading order after $A_4$ breaking. The flavon coupling to the charged leptons, $\varphi$, is a triplet of $A_4$ and its VEV should roughly preserve a $Z_3$ residual symmetry after $A_4$ breaking. $Z_3$ is phenomenologically necessary for realising TBM at leading order and has theoretically been realised in a lot of models. 
Depending on the different representation properties of $\varphi$, we consider two cases: i. $\varphi$ is a pseudo-real triplet of $A_4$, an economical case which introduces as few degrees of freedom as possible to the model, and ii. $\varphi$ is a complex triplet, a generalised case which can be regarded as a simplification of supersymmetric and multi-Higgs models. 
The breaking of $A_4 \to Z_3$ results in three physical parameters in case i: the scale of flavour symmetry breaking $v_\phi$ and flavon masses $m_{\varphi_1}$ and $m_{\varphi_2}$. In case ii, the mixing between two $Z_3$-covariant flavons introduces a mixing angle $\vartheta$, as well as one more flavon mass, which cannot be neglected. 
It is natural to assume the flavon masses to be of the same order of magnitude as $v_\varphi$. 

The only $Z_3$-preserving LFV processes are $\tau^- \to \mu^+ e^- e^-$, $\tau^- \to e^+ \mu^- \mu^-$.  They are triggered by the exchange of $Z_3$-covariant flavons and their branching ratios are dependent upon the flavour symmetry scale $v_\varphi$ and the flavon masses. In case i, both channels are mediated by the same flavon $\varphi_2$ and their branching ratios are approximately equal. 
In case ii, the two branching ratios are in general different from each other due to the presence of the mixing between the two $Z_3$-covariant flavons. If their mixing is maximal, the branching ratios are equal again. To be compatible with charged lepton masses, these processes in both cases are suppressed by ratios of charged lepton masses to the scale of flavour symmetry, to be exact, suppressed by $m_\mu m_\tau/v_\varphi^2$. Once we assume $v_\varphi$ around the electroweak scale, their branching ratios are much lower than current experimental limits. 

$Z_3$ is not an exact symmetry due to the interrupt with other fields. The breaking of $Z_3$ is also supported by the phenomenological requirement that TBM must gain corrections to match current oscillation data. In presence of $Z_3$ breaking, other 3-body LFV decays and all radiative decays forbidden by $Z_3$ can take place, but are suppressed by the small $Z_3$-breaking effects. 
We identify three $Z_3$-breaking effects: the mixing of charged lepton flavour eigenstates, the mixing between the $Z_3$-invariant and $Z_3$-covariant flavons, and the mass splitting of the two real degrees of freedom of each $Z_3$-covariant complex flavons. Under the assumption that the corrections to TBM come mainly from the $Z_3$-breaking effects, strong relations between mixing angles, especially $\theta_{13}$, with charged LFV processes arise. 

$Z_3$-breaking charged LFV processes depend on the explicit structure of a concrete model. We consider two models in cases i and ii, respectively. These models have been proposed in our former work \cite{PZ} and fit well with current oscillation data. 
We derive analytical expressions for branching ratios of all processes and numerically check the constraints on the $A_4$-breaking scale and flavon masses for each model. The results in Model I are quite simple. In special, a sum rule of $\tau^-\to \mu^+\mu^-e^-$, $\tau^-\to \mu^+\mu^-\mu^-$ and $\tau^-\to e^+ e^- \mu^- $ is obtained and all these processes are suppressed by $2s_{13}^2+(1-\sqrt{3}s_{12})^2$ compared with the $Z_3$-preserving ones. Another relation is $\text{Br}(\tau^- \to \mu^- \gamma) \approx \text{Br}(\tau^- \to e^- \gamma)$. 
The most stringent constraint is from the $\mu^-\to e^-\gamma$ measurement. The main contributions are the mixing of charged leptons and the mass splitting of the $Z_3$-covariant flavons induced by the breaking of the $Z_3$ symmetry. 
In Model I, the only unknown parameters contribute to this process are the scale $v_\varphi$ and the $Z_3$-covariant flavon mass $m_{\varphi_2}$. Their allowed parameter space can be derived from the current upper limit for this process. Setting $v_\varphi$ around the electroweak scale, we arrive at $m_{\varphi_2}>500$ GeV. 
The results in Model II are strongly dependent upon the mixing between the $Z_3$-covariant flavons. They match with those in Model I in the limit $\vartheta=\pm 45^\circ$ and one of the $Z_3$-covariant flavons decouple from the processes. 
In both models, branching ratios of $\tau^-\to e^+ e^-e^-$ and $\mu^-\to e^+ e^-e^-$ are much weaker than the other processes due to the suppression of electron mass. 
The flavon masses and their mixing angle $\vartheta$ will influence the branching ratio. Tuning these parameters, tiny masses,  300-400 GeV, for the $Z_3$-breaking flavons are allowed by experiment constraints. 

In conclusion, we have studied flavon-induced charged LFV processes. These flavons give explanation to lepton flavour mixing in flavour models and contribute to the other LFV processes beyond neutrino oscillations. Different from most models assuming the flavour symmetry at very high energy scale to avoid the strong constraints from charged LFV processes, we have checked that a relatively low-scale flavour symmetry, not far from the electroweak scale, is consistent with current experiment constraints from charged LFV processes. The main reason is that since these models give explanation to the flavour mixing, the couplings between flavons and charged leptons should be reasonably suppressed by the charged lepton masses. Furthermore, the approximative residual symmetry in the charged lepton sector, strongly suggested in most flavour models and supported by current oscillation data, can give an additional suppression factor $\mathcal{O}(0.01)$ for most LFV processes of 3-body charged lepton decays and all radiative decays. 
The method we developed here for calculating flavon-induced charged LFV processes can be extended into models with different flavour symmetries, such as $S_4$ and $A_5$.

\section*{Acknowledgement}

We would like to thank Yu-Feng Li, Richard Ruiz, Cedric Weiland and Shun Zhou for their useful discussions.  
This work was supported by the European Research Council under ERC Grant “NuMass” (FP7-IDEAS-ERC ERC-CG 617143), H2020 funded ELUSIVES ITN (H2020-MSCA-ITN-2015, GA-2015-674896-ELUSIVES) and InvisiblePlus (H2020-MSCA-RISE-2015, GA-2015-690575-InvisiblesPlus).

%%%%%%%%%%%%%%%%%%%%%%%%%%%%%%%%%%%%%%%%%%%%%%%%%%%%%%%%%%%%%%%%%%%%%%%%%%%%%%%%%%%%%%%%%%%%%%%%%%%%%%%%%%%%%%%%%%%%%%%%%%%%%%%%%%%%%%%%%%%

\appendix

\section{Radiative decays mediated by complex scalars}

Some of the scalars in our models (e.g., $\varphi_2$ in Model I, and $\varphi_2'$, $\varphi_3'$ in Model II) are complex scalars. Their couplings to fermions are strongly dependent upon the chirality of fermions. 
Thus, the chirality of fermions should be taken into account carefully when we calculate $l_1^-\to l_2^-\gamma$. 
It is useful for us to re-express $A_L^{l_2l_1} P_L+ A_R^{l_2l_1} P_R$ in Eq.~\eqref{eq:EDM} as
\begin{eqnarray}
A_L^{l_2l_1} P_L+ A_R^{l_2l_1} P_R = g_{LL} O_{LL}^{l_2l_1}(\varphi) + g_{RR} O_{RR}^{l_2l_1}(\varphi) + g_{RL} O_{RL}^{l_2l_1}(\varphi) +g_{LR} O_{LR}^{l_2l_1}(\varphi) \,,
\end{eqnarray}
where $O_{P_2P_1}^{l_2l_1}(\varphi)$ (for $P_1,P_2=L,R$) stand for the following loop integral excluding the $\sigma_{\mu\nu}q^\nu$ part: 
\begin{eqnarray}
\int\frac{d^4p}{(2\pi)^4} P_{P_2} \frac{i( \not p_2- \not p)+m_\tau}{(p_2-p)^2-m_\tau} \gamma_\mu \frac{i( \not p_1- \not p)+m_\tau}{(p_1-p)^2-m_\tau} P_{P_1} \frac{i}{p^2-m_\varphi^2}\,,
\end{eqnarray}
and $g_{P_2P_1}$ are the relevant coefficients. 
Neglecting subleading terms, we derive the following expression for $O_{P_1P_2}^{l_2l_1}(\varphi)$ as 
\begin{eqnarray}
&&O_{LL}^{l_2l_1}(\varphi)=\frac{1}{(4\pi)^2}\frac{m_\tau}{ m_\varphi^2}\big(\log \frac{m_\tau^2}{m_\varphi^2}+\frac{3}{2}\big) P_L \,,\nonumber\\
&&O_{RR}^{l_2l_1}(\varphi)=\frac{1}{(4\pi)^2}\frac{m_\tau}{ m_\varphi^2}\big(\log \frac{m_\tau^2}{m_\varphi^2}+\frac{3}{2}\big) P_R \,,\nonumber\\
&&O_{RL}^{l_2l_1}(\varphi)=\frac{1}{(4\pi)^2}\frac{m_{l_1}}{ m_\varphi^2}\times \frac{-1}{6} P_L \,,\nonumber\\
&&O_{LR}^{l_2l_1}(\varphi)=\frac{1}{(4\pi)^2}\frac{m_{l_1}}{ m_\varphi^2}\times \frac{-1}{6} P_R \,,
\label{eq:Integral}
\end{eqnarray}
For $\tau^- \to e^- \gamma, \mu^- \gamma$, $m_{l_1}=m_\tau$. For $\mu^-\to e^- \gamma$, $m_{l_1}=m_\mu$, and thus, $O_{LR}^{e\mu}, O_{RL}^{e\mu} \ll O_{LL}^{e\mu},O_{RR}^{e\mu}$. These results are compatible with \cite{radiative}. 

In Model I, we use the Lagrangian in Eqs.~\eqref{eq:Ltau2egamma}, \eqref{eq:Ltau2mugamma} and \eqref{eq:Lmu2egamma} to calculate the radiative decays. With the help of Eq.~\eqref{eq:Integral}, we derive
\begin{eqnarray}
A_L^{e\tau} P_L+ A_R^{e\tau} P_R &=& ie \tilde{\epsilon}_\varphi \frac{m_\tau^2}{ v_\varphi^2} \big[ O_{RR}^{e\tau}(\varphi_2) + O_{LR}^{e\tau}(\varphi_2) - O_{RR}^{e\tau}(\varphi_1) - O_{LR}^{e\tau}(\varphi_1) \big] \,,\nonumber\\
A_L^{\mu\tau} P_L+ A_R^{\mu\tau} P_R &=& ie \tilde{\epsilon}_\varphi^* \frac{ m_\tau^2}{v_\varphi^2} \big[ O_{RR}^{\mu\tau}(\varphi_2) + O_{LR}^{\mu\tau}(\varphi_2) - O_{RR}^{\mu\tau}(\varphi_1) - O_{LR}^{\mu\tau}(\varphi_1) \big] \,,\nonumber\\
A_L^{e\mu} P_L+ A_R^{e\mu} P_R &=& ie e^{-i\theta_{\varphi}} \frac{ m_\tau}{v_\varphi^2} \left\{ m_\mu \big[ O_{RR}^{e\mu}(h_2) - O_{RR}^{e\mu}(a_2) \right] + m_\tau \left[ O_{LR}^{e\mu}(h_2) - O_{LR}^{e\mu}(a_2) \big] \right\} 
\end{eqnarray}
for $\tau^-\to e^-\gamma$, $\tau^-\to\mu^-\gamma$ and $\mu^-\to e^- \gamma$, respectively. 
A direct calculation shows that $O_{RR}^{l_2\tau}(\varphi) + O_{LR}^{l_2\tau}(\varphi)=F(\varphi)/m_\tau P_R$ (for $l_2=e,\mu$) and $m_\mu O_{RR}^{e\mu}(\varphi) + m_\tau O_{LR}^{e\mu}(\varphi)= m_\mu F(\varphi)/m_\tau P_R$. 
And using the approximation 
\begin{eqnarray}
F(h_2)-F(a_2)=-4|\epsilon_\varphi|\frac{m_\tau^2}{m_{\varphi_2}^2} \big(\log\frac{m^2_\tau}{m_{\varphi_2}^2}+\frac{7}{3} \big)\,,
\end{eqnarray}
we finally obtain the results in Eq.~\eqref{eq:AR}. 

In Model II, The related Lagrangian terms for $\tau^-\to e^-\gamma$, $\tau^-\to \mu^-\gamma$ and $\mu^-\to e^-\gamma$ are respectively given by
\begin{eqnarray}
&&\frac{m_\tau}{v_\varphi} \overline{e_L} \tau_R 
\Big[-(1+\sqrt{2}c_{12}c_\vartheta+\sqrt{2}c_{13}s_\vartheta) \epsilon_\varphi h_1 + \sqrt{2}c_\vartheta \varphi_2' + \sqrt{2}s_\vartheta \varphi_3^{\prime*} \Big] \nonumber\\
 %%%%
&+&\frac{m_\tau}{v_\varphi} \overline{\tau_L} \tau_R 
\Big[h_1 + (c_{12}-\sqrt{2}s_\vartheta) \epsilon_\varphi \varphi_2^{\prime*} + (c_{13}+\sqrt{2}c_\vartheta ) \epsilon_\varphi \varphi_3' \Big] \nonumber\\
 %%%%
&+&\frac{m_\tau}{v_\varphi} \overline{\tau_R} \tau_L 
\Big[h_1 + (c_{12}+\sqrt{2}c_\vartheta) \epsilon_\varphi \varphi_2^{\prime*} + (c_{13}+\sqrt{2}s_\vartheta ) \epsilon_\varphi \varphi_3' \Big]  \,,\nonumber\\
%%%%%%%%%%%%%%%%%%%%%
&&\frac{m_\tau}{v_\varphi} \overline{\mu_L} \tau_R 
\Big[-(1-\sqrt{2}c_{12}s_\vartheta+\sqrt{2}c_{13}c_\vartheta) \epsilon_\varphi^* h_1 - \sqrt{2}s_\vartheta \varphi_2^{\prime*} + \sqrt{2}c_\vartheta \varphi_3' \Big] \nonumber\\
 %%%%
&+&\frac{m_\tau}{v_\varphi} \overline{\tau_L} \tau_R 
\Big[h_1 + (c_{12}+\sqrt{2}c_\vartheta) \epsilon_\varphi^* \varphi_2' + (c_{13}+\sqrt{2}s_\vartheta ) \epsilon_\varphi^* \varphi_3^{\prime*} \Big] \nonumber\\
 %%%%
&+&\frac{m_\tau}{v_\varphi} \overline{\tau_R} \tau_L 
\Big[h_1 + (c_{12}-\sqrt{2}s_\vartheta) \epsilon_\varphi^* \varphi_2' + (c_{13}+\sqrt{2}c_\vartheta ) \epsilon_\varphi^* \varphi_3^{\prime*} \Big]  \,,\nonumber\\
%%%%%%%%%%%%%%%%%%%%%
&&e^{-i\frac{\theta_\varphi}{2}}\frac{m_\tau}{v_\varphi} \overline{e_L} \tau_R 
\Big[\left( c_\vartheta + (1-\kappa) |\epsilon_\varphi| s_\vartheta \right) h_2' + i\left(c_\vartheta -  (1-\kappa) |\epsilon_\varphi| s_\vartheta \right)  a_2' \nonumber\\
&&\hspace{2.2cm}+ 
\left( s_\vartheta -  (1-\kappa) |\epsilon_\varphi| c_\vartheta \right) h_3' + i\left(s_\vartheta +  (1-\kappa) |\epsilon_\varphi| c_\vartheta \right)  a_3'
 \Big] \nonumber\\
 %%%%
&+&e^{-i\frac{\theta_\varphi}{2}}\frac{m_\tau}{v_\varphi} \overline{\tau_R} \tau_L 
\Big[\left( -s_\vartheta +  (1+\kappa) |\epsilon_\varphi| c_\vartheta \right) h_2' - i\left(s_\vartheta +  (1+\kappa) |\epsilon_\varphi| c_\vartheta \right)  a_2' \nonumber\\
&&\hspace{2.2cm}+ 
\left( c_\vartheta +  (1+\kappa) |\epsilon_\varphi| s_\vartheta \right) h_3' + i\left(c_\vartheta -  (1+\kappa) |\epsilon_\varphi| s_\vartheta \right)  a_3' 
 \Big] \nonumber\\
 %%%%
&+& e^{-i\frac{\theta_\varphi}{2}}\frac{m_\mu}{v_\varphi} \overline{\tau_L} \tau_R 
\Big[\left( c_\vartheta -  (1-\kappa) |\epsilon_\varphi| s_\vartheta \right) h_2' + i\left(c_\vartheta +  (1-\kappa) |\epsilon_\varphi| s_\vartheta \right)  a_2' \nonumber\\
&&\hspace{2.2cm}+  
\left( s_\vartheta +  (1-\kappa) |\epsilon_\varphi| c_\vartheta \right) h_3' + i\left(s_\vartheta -  (1-\kappa) |\epsilon_\varphi| c_\vartheta \right)  a_3'
 \Big]  \,.
\label{eq:Feynman_radiative2}
\end{eqnarray}
They results in expressions of $A_L^{l_2l_1} P_L+A_R^{l_2l_1} P_R$ given by 
\begin{eqnarray}
 ie \epsilon_\varphi \frac{m_\tau^2}{ v_\varphi^2}& \Big\{& -(1+\kappa+\sqrt{2}c_\vartheta c_{12}+\sqrt{2}s_\vartheta c_{13}) \big[O_{RR}^{e\tau}(h_1) + O_{LR}^{e\tau}(h_1)\big] \nonumber\\
&&+ \sqrt{2} c_\vartheta c_{12} \big[ O_{RR}^{e\tau}(\varphi_2) + O_{LR}^{e\tau}(\varphi_2)\big]
- 2 s_\vartheta c_\vartheta (1+\kappa) O_{RR}^{e\tau}(\varphi_2) + 2 c_\vartheta^2 (1-\kappa) O_{LR}^{e\tau}(\varphi_2)\big] \nonumber\\
&&+ \sqrt{2} s_\vartheta c_{13} \big[ O_{RR}^{e\tau}(\varphi_3) + O_{LR}^{e\tau}(\varphi_3)\big]
+ 2 s_\vartheta c_\vartheta (1+\kappa) O_{RR}^{e\tau}(\tau,\varphi_3) + 2 s_\vartheta^2 (1-\kappa) O_{LR}^{e\tau}(\varphi_3)\big] \Big\} \,,\nonumber\\
 ie \epsilon_\varphi^* \frac{m_\tau^2}{ v_\varphi^2} &\Big\{& -(1-\kappa-\sqrt{2}s_\vartheta c_{12}+\sqrt{2}c_\vartheta c_{13}) \big[O_{RR}^{\mu\tau}(h_1) + O_{LR}^{\mu\tau}(h_1)\big] \nonumber\\
&&- \sqrt{2} s_\vartheta c_{12} \big[ O_{RR}^{\mu\tau}(\varphi_2) + O_{LR}^{\mu\tau}(\varphi_2)\big]
- 2 s_\vartheta c_\vartheta (1-\kappa) O_{RR}^{\mu\tau}(\varphi_2) + 2 s_\vartheta^2 (1+\kappa) O_{LR}^{\mu\tau}(\varphi_2)\big] \nonumber\\
&&+ \sqrt{2} c_\vartheta c_{13} \big[ O_{RR}^{\mu\tau}(\varphi_3) + O_{LR}^{\mu\tau}(\varphi_3)\big]
+ 2 s_\vartheta c_\vartheta (1-\kappa) O_{RR}^{\mu\tau}(\varphi_3) + 2 c_\vartheta^2 (1+\kappa) O_{LR}^{\mu\tau}(\varphi_3)\big] \Big\} \,,\nonumber\\
 ie \frac{ m_\tau}{v_\varphi^2} &\Big\{& e^{-i\theta_{\varphi}} m_\mu c_\vartheta^2 \big[ O_{RR}^{e\mu}(h_2) - O_{RR}^{e\mu}(a_2) \big] - e^{-i\theta_{\varphi}} m_\tau c_\vartheta s_\vartheta \big[ O_{LR}^{e\mu}(h_2) - O_{LR}^{e\mu}(a_2) \big] \nonumber\\
&& +e^{-i\theta_{\varphi}} m_\mu s_\vartheta^2 \big[ O_{RR}^{e\mu}(h_3) - O_{RR}^{e\mu}(a_3) \big] 
+ e^{-i\theta_{\varphi}} m_\tau c_\vartheta s_\vartheta \big[ O_{LR}^{e\mu}(h_3) + O_{LR}^{e\mu}(a_3) \big] \nonumber\\
&& + \epsilon_\varphi^* m_\tau (c_\vartheta^2-s_\vartheta^2+\kappa) \big[ O_{LR}^{e\mu}(h_2) + O_{LR}^{e\mu}(a_2) - O_{LR}^{e\mu}(h_3) - O_{LR}^{e\mu}(a_3) \big] \Big\}
\label{eq:AR22}
\end{eqnarray}
for $\tau^-\to e^-\gamma$, $\tau^-\to\mu^-\gamma$ and $\mu^-\to e^- \gamma$, respectively. We express $O_{LR}^{l_2l_1}(\varphi)$ and $O_{RR}^{l_2l_1}(\varphi)$ in $F(\varphi)$, expand $m_{h_2'}$, $m_{a_2'}$ and $m_{h_3'}$, $m_{a_3'}$ around $m_{\varphi_2'}$ and $m_{\varphi_3'}$, respectively, and finally, we obtain Eq.~\eqref{eq:AR2}.

Compared with $A_R^{l_2l_1}$, $A_L^{l_2l_1}$ is negligibly small in all three channels. The reason is that $e_R$, $\mu_R$ and $\tau_R$ are singlets of $A_4$. Each column of the charged lepton mass matrix in Eq. \eqref{eq:mass_shift1} is proportional to one charged lepton mass. This kind of flavour structure leads to very small mixing of $e_R$, $\mu_R$ and $\tau_R$ (suppressed by both $\epsilon_\varphi$ and charged lepton mass ratio), and thus terms such as $\overline{e}_L\tau_R \varphi_i$ are highly suppressed. The feature $A_L^{l_2l_1}\ll A_R^{l_2l_1}$ is a feature in models where right-handed charged leptons belong to singlet representations of the flavour symmetry. Considerable mixing among right-handed charged leptons is possible in some other models. In that case, the contribution of $A_L^{l_2l_1}$ should be included in the decay $l_1^-\to l_2^- \gamma$.

\end{document}